\documentclass[referee]{raa}            
\usepackage{graphicx,times}             

\begin{document}

   \title{Long-term $BVRI$ photometric light curves of 15 PMS stars in\\ the IC 5070 star-forming region
$^*$
\footnotetext{$*$ Based on observations obtained at the Rozhen National Astronomical Observatory, Bulgaria. The photometric data are only available at the CDS.}
}

   \volnopage{Vol.0 (20xx) No.0, 000--000}      
   \setcounter{page}{1}          

   \author{Sunay Ibryamov
      \inst{1}
   \and Evgeni Semkov
      \inst{2}
   \and Teodor Milanov
      \inst{1}
	 \and Stoyanka Peneva
      \inst{2}
   }

   \institute{Department of Physics and Astronomy, Faculty of Natural Sciences, University of Shumen,
             115, Universitetska Str., 9700 Shumen, Bulgaria; {\it sibryamov@shu.bg}\\
        \and
             Institute of Astronomy and National Astronomical Observatory, Bulgarian Academy of Sciences, 72, Tsarigradsko Shose Blvd., 1784 Sofia, Bulgaria\\
   }

   \date{Received~~...; accepted~~...}

\abstract{
This paper reports the results from the multicolor photometric observations of 15 pre-main sequence stars collected in the period 2010 September $-$ 2017 October.
The stars from our study are located in the star-forming HII region IC 5070.
These objects were previously detected as either emission line stars, flare stars, T Tauri variables or Herbig Ae/Be stars. 
Photometric observations, especially concerning the long-term behavior of the objects are missing in the literature.
We present the first photometric monitoring for all stars from our study.
The analysis of the obtained $BVRI$ photometric data allows to draw a conclusion that all investigated objects are variable stars.
In the case of LkH$_\alpha$ 146 we identified previously unknown periodicity in its photometric variability.
}

   \authorrunning{S. Ibryamov, E. Semkov, T. Milanov \& S. Peneva}            
   \titlerunning{Long-term $BVRI$ light curves of 15 PMS stars in the Pelican Nebula}  

   \maketitle

%
%
\section{Introduction}           
\label{sect:intro}

The North-America and Pelican nebulae (NGC 7000/IC 5070) represent one of the most notable and well-studied active star-forming complexes, where a large number of young pre-main sequence (PMS) stars, cometary nebulae, collimated jets and Herbig-Haro objects can be found (Herbig~\cite{herb58}, Guieu et al.~\cite{guie09}, Rebull et al.~\cite{rebu11} and Bally et al.~\cite{ball14}).
Recent major studies of the NGC 7000/IC 5070 complex were made by Rebull et al.~(\cite{rebu11}), Armond et al.~(\cite{armo11}), Zhang et al.~(\cite{zhan14}), Bally et al.~(\cite{ball14}) and Damiani et al.~(\cite{dami17}).

\newpage
One of the most important features of PMS stars is their photometric and spectroscopic variability discovered at the beginning of their study.
PMS stars are separated into two types $-$ the low-mass T Tauri stars (TTSs) and the more massive Herbig Ae/Be stars (HAEBESs).
The TTSs exhibit strong irregular photometric variability and emission spectra.
TTSs are divided into two sub-classes $-$ classical T Tauri stars (CTTSs), still actively accreting from their circumstellar disks, and the weak-line T Tauri stars (WTTSs) which show no signs of disk accretion (M\'{e}nard \& Bertout~\cite{mena99}).

The superpositions of cool and hot spots on the stellar surface; flare-like events; variable mass accretion rate from the circumstellar disk onto the stellar surface; as well as circumstellar dust or clouds obscuration events are the possible causes for the observed variability of the TTSs (Grinin et al.~\cite{grin91}, Herbst et al.~\cite{herb94}, Ismailov~\cite{isma05} and Herbst et al.~\cite{herb07}).
The review of various classification schemes of photometric variability of young stars can be found in the work of Cody et al.~(\cite{cody14}).
Especially for CTTSs Ismailov~(\cite{isma05}) proposed classification scheme based on the light curve shape.

Unlike TTSs, the HAEBESs are less photometrically active and less studied.
Detailed descriptions of the observed features of HAEBESs are given in Hillenbrand et al.~(\cite{hill92}), Perez \& Grady~(\cite{pere97}) and Waters \& Waelkens~(\cite{wate98}).
Evidence of stellar winds, jets and mass accretion in the HAEBESs was not found (Waters \& Waelkens~\cite{wate98}).
The variability (if any) of these stars derives from cool spots and/or from the transit of disc clumps.
The obscuration events are very likely to be present in most HAEBESs but they can be only registered when the circumstellar disks are located at a small angle to the line of sight (Grinin et al.~\cite{grin91} and Natta \& Whitney~\cite{natt00}).
When PMS stars approach the main sequence they lose their distinctive features and at this stage they are hardly different from main sequence stars.

The stars from our study were selected from the SIMBAD database by exact object types (Variable star of Orion type, Emission line-star, Flare star, and Herbig Ae/Be star) and with the condition on their location $-$ position within 20 arcmin around the well-studied young stellar object V2492 Cyg (see Covey et al.~\cite{cove11}, Aspin~\cite{aspi11}, K\'{o}sp\'{a}l et al.~\cite{kosp13}, Hillenbrand et al.~\cite{hill13}, Giannini et al.~\cite{gian18} and Ibryamov et al.~\cite{ibry18}).
Photometric observations, especially concerning the long-term behavior of the stars from our study are missing in the literature.
Long-term photometric observations are important for the exact classification of PMS stars.
Such observations are directed at the active star-formation fields with the goal of finding and classifying the variability of the young stellar objects (YSOs) embedded in them.

The present paper is a part of our program for the photometric study of PMS stars located in the NGC 7000/IC 5070 star-forming complex.
The results from our recent studies have been published in Ibryamov et al.~(\cite{ibry15}), Ibryamov \& Semkov~(\cite{ibry16}), Semkov et al.~(\cite{semk17}) and Ibryamov et al.~(\cite{ibry18}).
Section 2 in the present paper gives information about the process of acquiring photometric observations and data reduction.
Section 3 describes the obtained results and their analysis.

\newpage
\section{Observations and data reduction}
\label{sect:Obs}

The CCD observations reported in the paper were collected during the time period from 2010 September to 2017 October.
All CCD observations were obtained with two telescopes $-$ the 50/70-cm Schmidt and the 60-cm Cassegrain $-$ administered by the Rozhen National Astronomical Observatory in Bulgaria.
The number of observational nights used to estimate the brightness of each object is 119.

The observations were performed with two different types of CCD cameras $-$ FLI PL16803 (4096 $\times$ 4096 pixels and 9 $\times$ 9 $\mu m/$pixel size) on the 50/70-cm Schmidt telescope and FLI PL09000 (3056 $\times$ 3056 pixels and 12 $\times$ 12 $\mu m/$pixel size) on the 60-cm Cassegrain telescope.
All frames were taken through a standard Johnson$-$Cousins ($BVRI$) set of filters.
The frames are dark frame subtracted and flat field corrected.
The photometric data were reduced using \textsc{idl} based \textsc{daophot} subroutine.
As a reference, the $BVRI$ comparison sequence of eleven stars in the field around V2492 Cyg reported in Ibryamov et al.~(\cite{ibry18}) was used.
All data were analyzed using the same aperture, which was chosen to have a 4 arcsec radius, while the background annulus was taken from 9 to 14 arcsec.
The mean value of the errors in the reported magnitudes is 0.01$-$0.02 mag for the $I$- and $R$-band data and 0.01$-$0.03 mag for the $V$- and $B$-band data.

\section{Results and discussion}
\label{sect:Res}

\begin{figure*}
   \centering
   \includegraphics[width=12cm, angle=0]{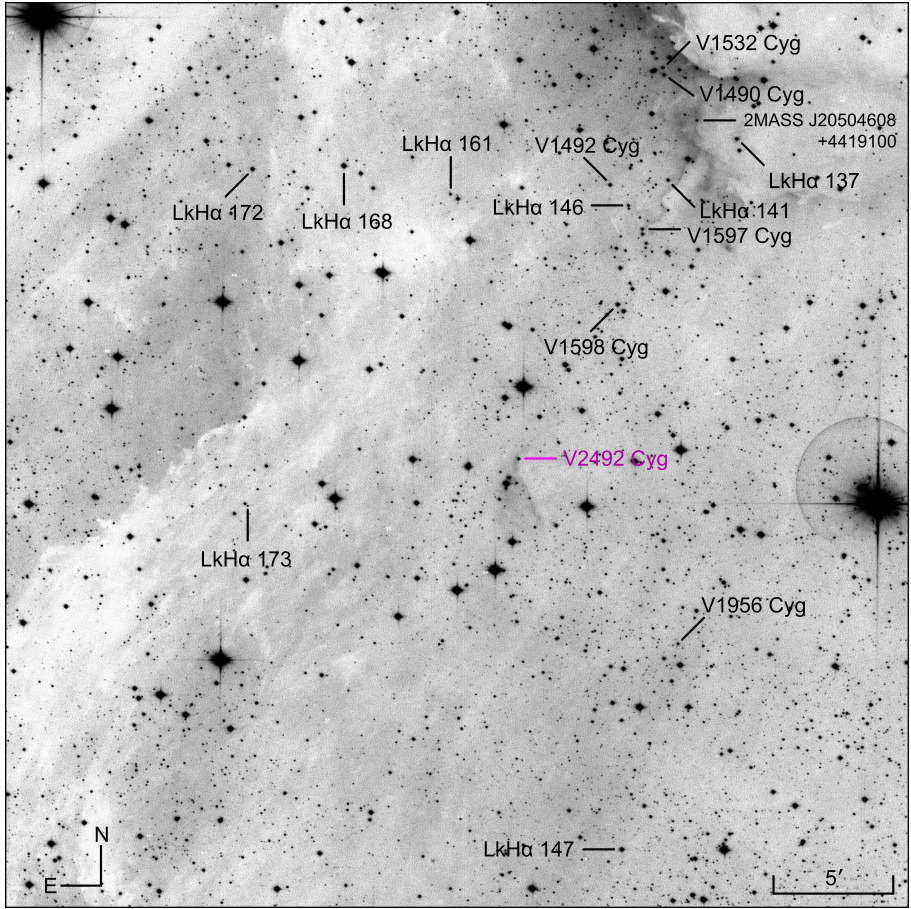}
   \caption{An image of the field around V2492 Cyg obtained on 2013 September 5 in $R$-band with the 50/70-cm Schmidt telescope at Rozhen NAO. The stars from our study and V2492 Cyg are marked.}\label{Fig:field}
   \end{figure*}

The stars from our study are listed in Table~\ref{Tab:designations} in the order of rising right ascension.
Star identifiers used in this paper are marked in boldface.
Fig.~\ref{Fig:field} shows an image of the field around V2492 Cyg, where the positions of the objects are marked.
The registered during our photometric monitoring minimal and maximal magnitudes and the amplitudes of variability in the $BVRI$-bands of the stars are given in Table~\ref{Tab:amplitudes}.

{\footnotesize
\begin{table*}[]
  \caption{Designations and coordinates of the stars from our study.}\label{Tab:designations}
  \begin{center}
  \begin{tabular}{llcrcccc}
	  \hline\hline
	  \noalign{\smallskip}
Nr  & GCVS$^1$ & HBC$^2$ & [KW97]$^3$ & LkH$_{\alpha}$ $^4$ & 2MASS ID$^5$ & RA$_{J2000.0}$ & Dec$_{J2000.0}$\\
   \noalign{\smallskip}
   \hline
   \noalign{\smallskip}
1  &                    & 698 & 50-12 & \textbf{137} & J20503703+4418247 & 20 50 37.03 & +44 18 24.7 \\
2  &                    &     &       &     & \textbf{J20504608+4419100} & 20 50 46.08 & +44 19 10.1 \\
3  & \textbf{V1956 Cyg} &     &       &              & J20505014+4357536 & 20 50 50.14 & +43 57 53.7 \\
4  &                    & 702 & 50-22 & \textbf{141} & J20505257+4416441 & 20 50 52.57 & +44 16 44.2 \\
5  & \textbf{V1490 Cyg} &     &       &              & J20505357+4421008 & 20 50 53.58 & +44 21 00.9 \\
6  & \textbf{V1532 Cyg} & 703 &       & 143          & J20505378+4421185 & 20 50 53.78 & +44 21 18.5 \\
7  & \textbf{V1597 Cyg} &     &       &              & J20505838+4414444 & 20 50 58.38 & +44 14 44.4 \\
8  &                    & 704 & 50-29 & \textbf{146} & J20510157+4415420 & 20 51 01.57 & +44 15 42.1 \\
9  &                    & 705 & 50-30 & \textbf{147} & J20510271+4349318 & 20 51 02.71 & +43 49 31.9 \\
10 & \textbf{V1598 Cyg} &     &       &              & J20510393+4411406 & 20 51 03.93 & +44 11 40.6 \\
11 & \textbf{V1492 Cyg} &     &       &              & J20510570+4416322 & 20 51 05.70 & +44 16 32.3 \\
12 &                    & 714 & 50-53 & \textbf{161} & J20514191+4416082 & 20 51 41.91 & +44 16 08.2 \\
13 &                    & 717 & 51-4  & \textbf{168} & J20520604+4417160 & 20 52 06.05 & +44 17 16.1 \\
14 &                    & 298 & 51-15 & \textbf{172} & J20522676+4417066 & 20 52 26.76 & +44 17 06.6 \\
15 &                    &     & 51-16 & \textbf{173} & J20522740+4403259 & 20 52 27.40 & +44 03 25.9 \\
   	  \hline \hline
  \end{tabular}
  \end{center}
  {\textbf{References:} $^1$General Catalogue of Variable Stars (Samus et al.~\cite{samu17}); $^2$Herbig \& Bell~(\cite{herb88}); $^3$Kohoutek \& Wehmeyer~(\cite{koho97}); $^4$Herbig~(\cite{herb58}); $^5$2Micron All-Sky Survey (Skrutskie et al.~\cite{skru06})\\
	\textbf{Note:} Star identifiers used in this paper are marked in boldface}.
	\end{table*}} 

	\begin{table*}[]
	{\footnotesize
	\caption{The registered minimal and maximal magnitudes and the amplitudes of variability in the $BVRI$-bands of the stars from our study.}\label{Tab:amplitudes}
  \begin{center}
  \begin{tabular}{llcccccccccccc}
	  \hline\hline
	  \noalign{\smallskip}
Nr & Star & $B_{min}$ & $B_{max}$ & $V_{min}$ & $V_{max}$ & $R_{min}$ & $R_{max}$ & $I_{min}$ & $I_{max}$ & $\Delta{B}$ & $\Delta{V}$ & $\Delta{R}$ & $\Delta{I}$ \\ 		
   \noalign{\smallskip}
   \hline
   \noalign{\smallskip}
1  & LkH$_{\alpha}$ 137 & 18.53 & 17.13 & 16.64 & 15.46 & 15.26 & 14.25 & 13.65 & 12.95 & 1.40 & 1.18 & 1.01 & 0.70 \\
2  & J2050+4419         & 19.35 & 18.86 & 17.64 & 17.24 & 16.35 & 15.94 & 14.83 & 14.48 & 0.49 & 0.40 & 0.41 & 0.35 \\
3  & V1956 Cyg          & 17.46 & 17.25 & 16.26 & 16.09 & 15.45 & 15.25 & 14.47 & 14.27 & 0.21 & 0.17 & 0.20 & 0.20 \\
4  & LkH$_{\alpha}$ 141 & 17.34 & 17.02 & 15.79 & 15.46 & 14.90 & 14.46 & 13.93 & 13.46 & 0.32 & 0.33 & 0.44 & 0.47 \\
5  & V1490 Cyg          & 18.14 & 16.99 & 16.84 & 15.40 & 15.69 & 14.40 & 14.53 & 13.29 & 1.15 & 1.44 & 1.29 & 1.24 \\
6  & V1532 Cyg          & 18.56 & 17.78 & 16.80 & 16.14 & 15.44 & 14.91 & 13.87 & 13.51 & 0.78 & 0.66 & 0.53 & 0.36 \\
7  & V1597 Cyg          & 18.07 & 17.71 & 16.34 & 15.99 & 15.17 & 14.76 & 13.65 & 13.29 & 0.36 & 0.35 & 0.41 & 0.36 \\
8  & LkH$_{\alpha}$ 146 & 18.32 & 17.58 & 16.67 & 15.91 & 15.58 & 14.78 & 14.37 & 13.68 & 0.74 & 0.76 & 0.80 & 0.69 \\
9  &LkH$_{\alpha}$ 147  & 15.99 & 15.84 & 14.56 & 14.35 & 13.47 & 13.29 & 12.21 & 12.04 & 0.15 & 0.21 & 0.18 & 0.17 \\
10 & V1598 Cyg       & 15.27 & 14.97 & 14.28 & 14.01 & 13.69 & 13.43 & 13.13 & 12.87 & 0.30 & 0.27 & 0.26 & 0.26 \\
11 & V1492 Cyg       & 17.85 & 16.90 & 16.93 & 15.31 & 15.68 & 14.16 & 14.59 & 13.14 & 0.95 & 1.62 & 1.52 & 1.45 \\
12 & LkH$_{\alpha}$ 161 & 18.63 & 17.97 & 17.13 & 16.48 & 15.89 & 15.39 & 14.69 & 14.29 & 0.66 & 0.65 & 0.50 & 0.40 \\
13 & LkH$_{\alpha}$ 168 & 14.87 & 14.75 & 13.60 & 13.44 & 12.78 & 12.57 & 11.81 & 11.68 & 0.12 & 0.16 & 0.21 & 0.13 \\
14 & LkH$_{\alpha}$ 172 & 16.19 & 16.02 & 14.76 & 14.54 & 13.82 & 13.59 & 12.83 & 12.69 & 0.17 & 0.22 & 0.23 & 0.14 \\
15 & LkH$_{\alpha}$ 173 & -     & -     & 17.61 & 17.19 & 17.00 & 16.62 & 13.46 & 13.30 & -    & 0.42 & 0.38 & 0.16 \\
   	  \hline \hline
  \end{tabular}
  \end{center}}
  \end{table*}
	
\begin{figure}
\begin{center}
\includegraphics[width=7cm]{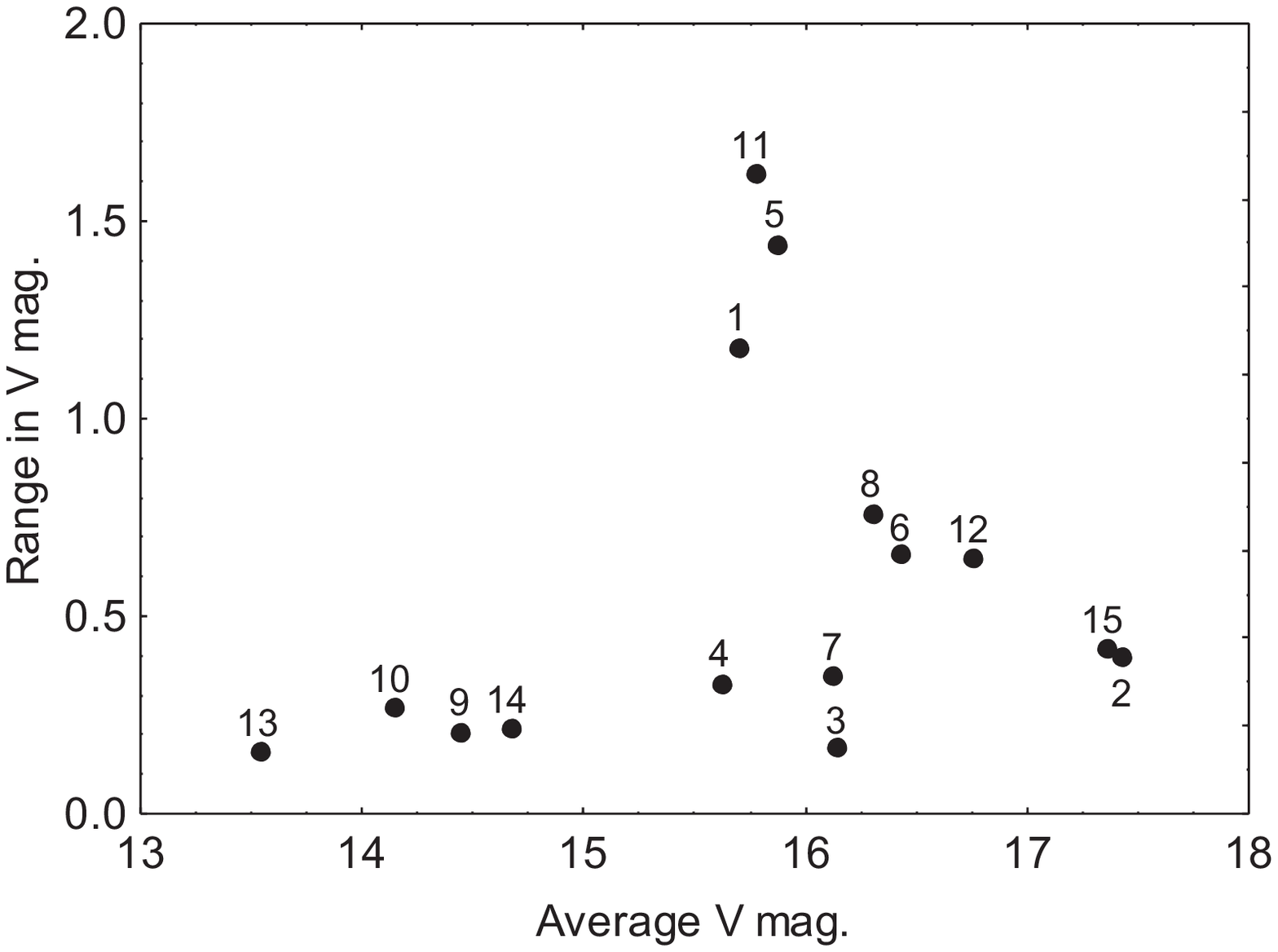}
\includegraphics[width=7cm]{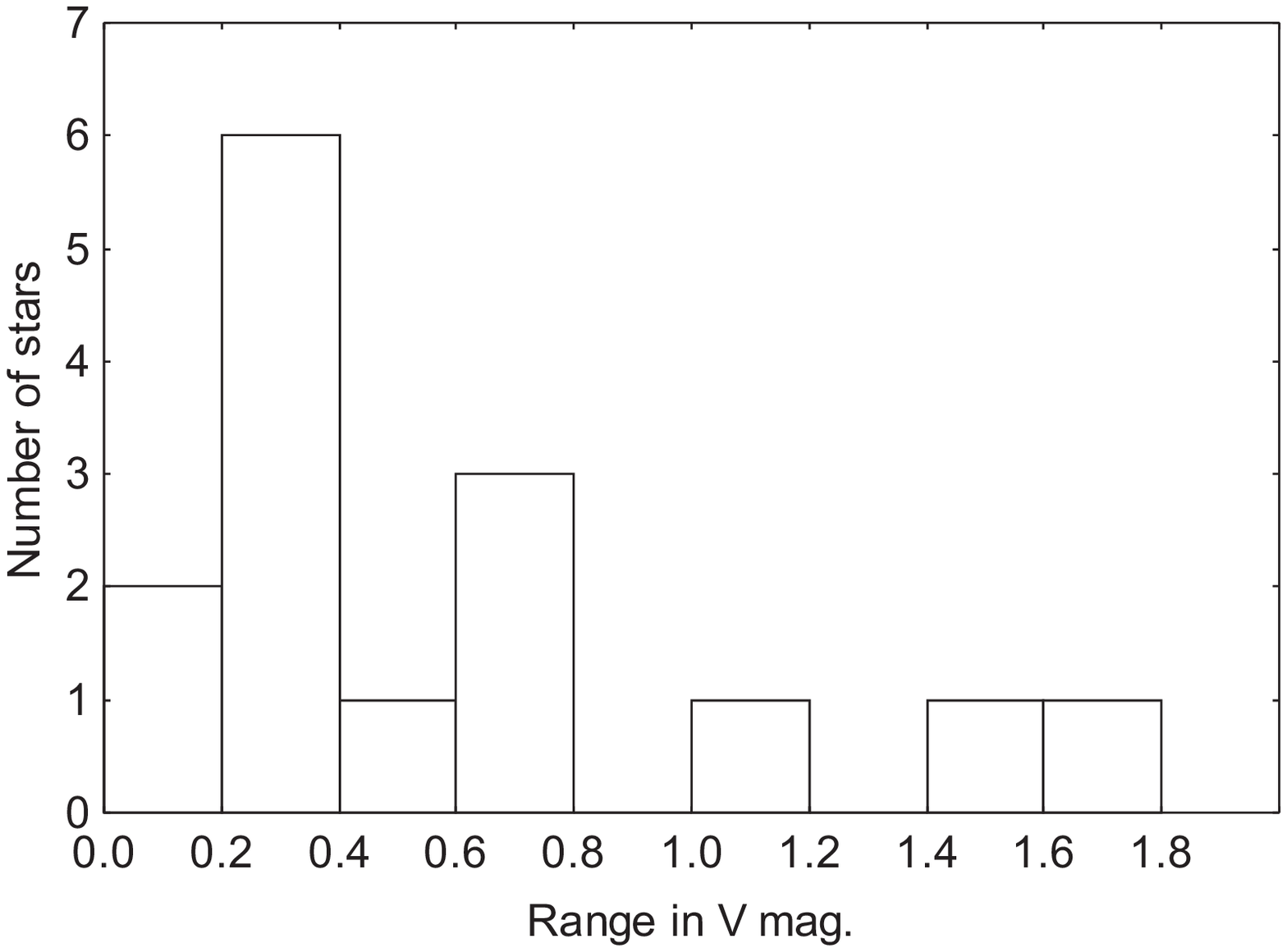}
\caption{Left: Range in $V$ versus average $V$ mag for stars from our study; Right: Histogram of $V$ mag ranges for stars from our study.}\label{Fig:hist}
\end{center}
\end{figure}

The $V$ magnitude range versus average brightness in the $V$-band of the investigated objects is shown in Fig.~\ref{Fig:hist} (left).
The used object designations are as in Table~\ref{Tab:designations}.
It is seen from the figure, that there is no apparent correlation between range and average brightness among the objects.
The histogram of $V$ magnitude ranges for the objects is shown in Fig.~\ref{Fig:hist} (right).
Roughly all objects (12) exhibit amplitude of variability on the long-term in the range between 0.1 and 0.8 mag.
The lowest amplitude star is LkH$_{\alpha}$ 168 ($\Delta V$= 0.16 mag).
Only 3 stars have an amplitude of variability larger than 1 mag.
They are LkH$_{\alpha}$ 137, V1490 Cyg and V1492 Cyg.
The largest amplitude star is V1492 Cyg ($\Delta V$= 1.62 mag).

\begin{figure}
   \centering
   \includegraphics[width=9cm, angle=0]{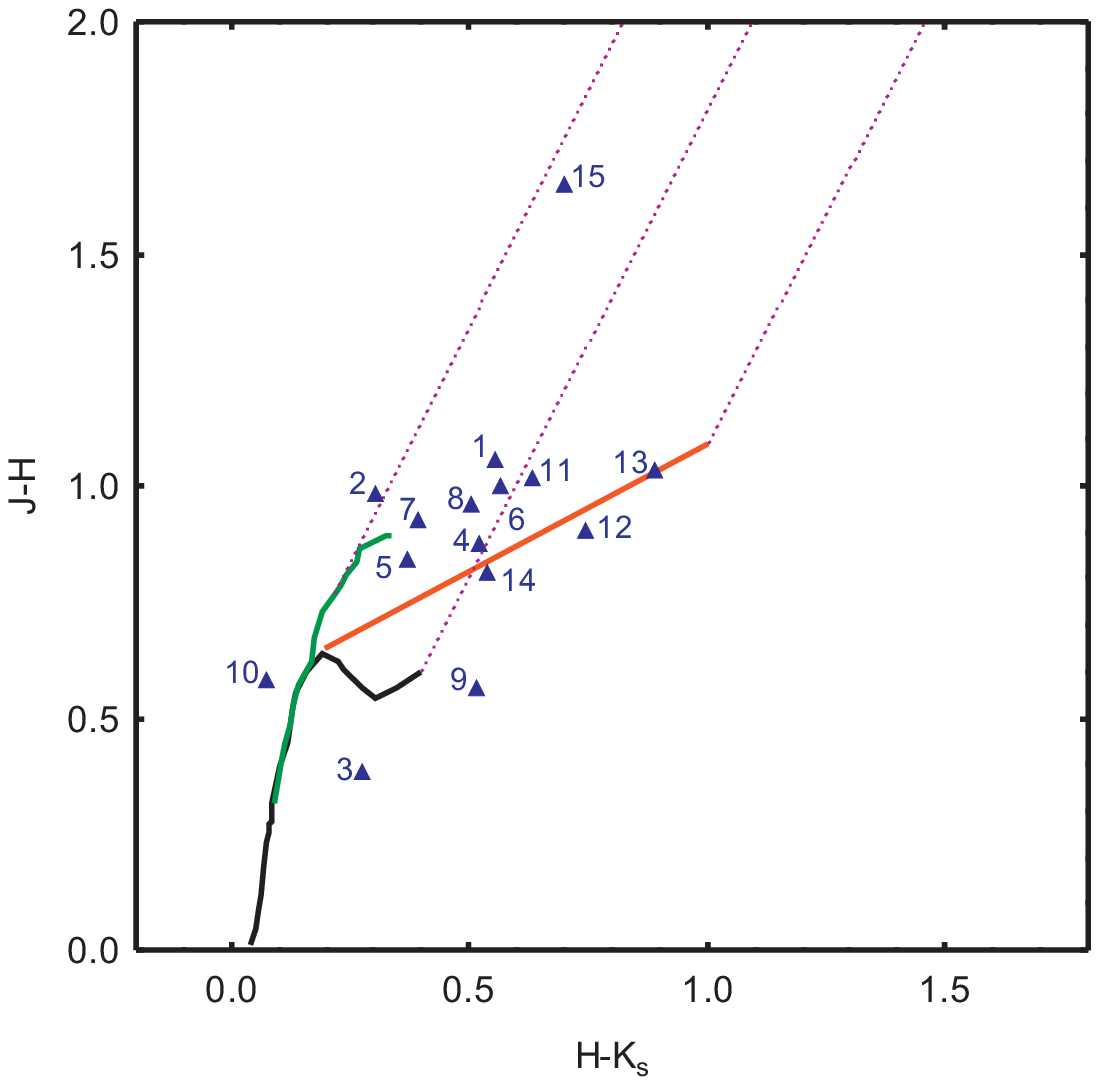}
   \caption{The $J-H$ versus $H-K_{s}$ diagram for the stars from our study detected in the three bands in 2MASS catalogue.}\label{Fig:2mass}
   \end{figure}

We used the 2MASS ($JHK_{s}$) magnitudes of the stars from our study to construct the two$-$color diagram ($J-H$/$H-K_{s}$) to identify the stars with infrared excess, indicating the presence of disks around them.
Fig.~\ref{Fig:2mass} shows the location of the main sequence (the dark line) and the giant stars (the green line) from Bessell \& Brett~(\cite{bess88}), and the CTTSs location (the orange line) from Meyer et al.~(\cite{meye97}).
A correction to the 2MASS photometric system was performed following the prescription of Carpenter~(\cite{carp01}).
The three parallel dotted lines show the direction of the interstellar reddening vectors determined for the NGC 7000/IC 5070 star-forming complex by Straiz\v{y}s et al.~(\cite{stra08}).
On Fig.~\ref{Fig:2mass} the objects are designated using their numbers given in Table~\ref{Tab:designations}.

\newpage	
Using data from our multicolor photometry we constructed three color$-$magnitude diagrams ($B-V$/$V$, $V-R$/$V$ and $V-I$/$V$) of the objects, which are displayed in Fig.~\ref{Fig:colors}.
For all objects we made time-series analysis for a periodicity search with the software package \textsc{period04} (Lenz \& Breger~\cite{lenz05}).
In the present paper we discuss the stars in groups with close photometric behaviors and features.

\subsection{LkH$_{\alpha}$ 137, V1490 Cyg, LkH$_{\alpha}$ 146 and V1492 Cyg}

Unlike the other stars studied in the paper, during our photometric monitoring LkH$_{\alpha}$ 137, V1490 Cyg, LkH$_{\alpha}$ 146 and V1492 Cyg show bigger amplitudes of the photometric variability.
These objects are included in the list of candidates for YSOs published by Guieu et al.~(\cite{guie09}).

\begin{figure}
\begin{center}
\includegraphics[width=7cm]{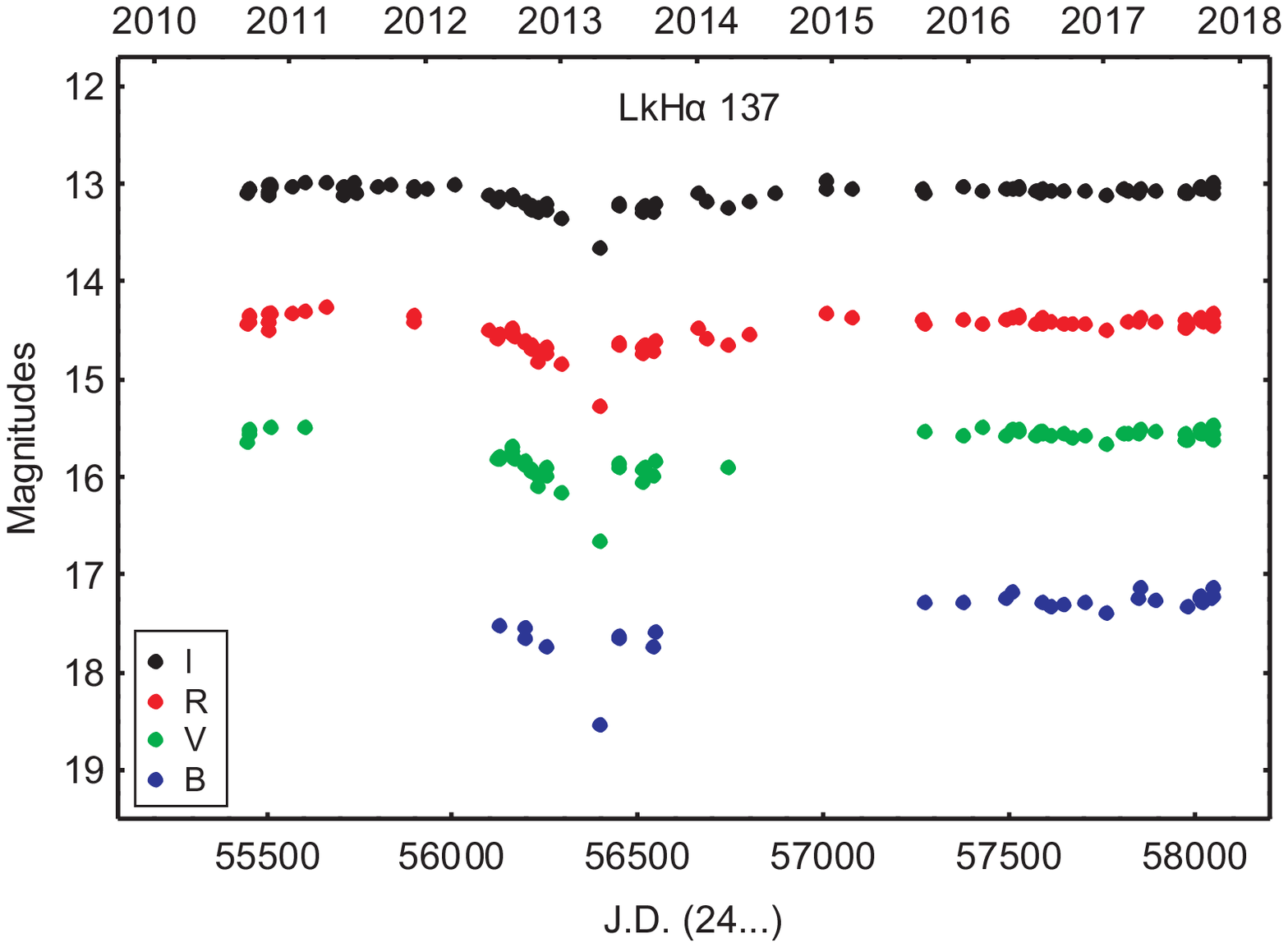}
\includegraphics[width=7cm]{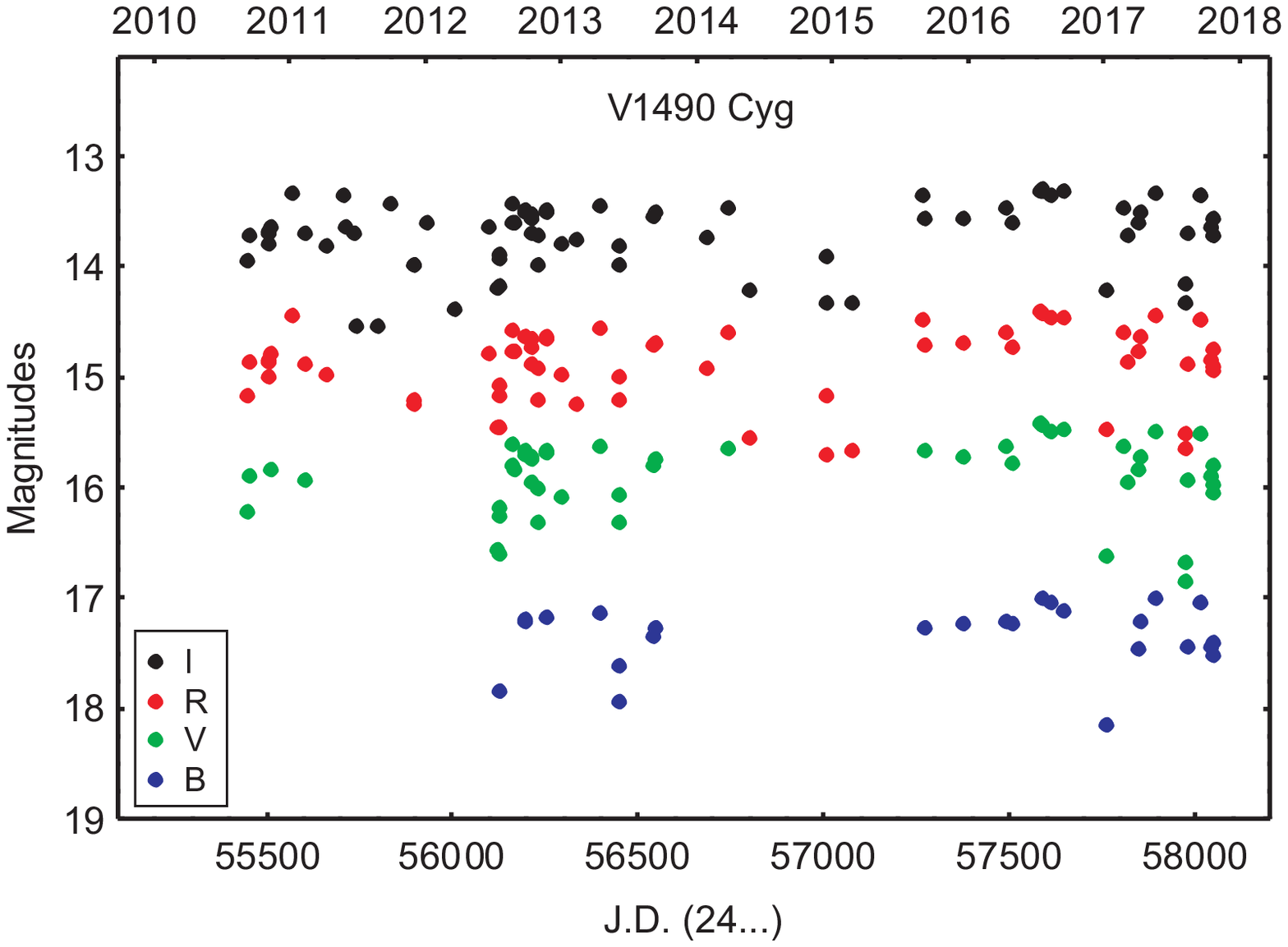}
\includegraphics[width=7cm]{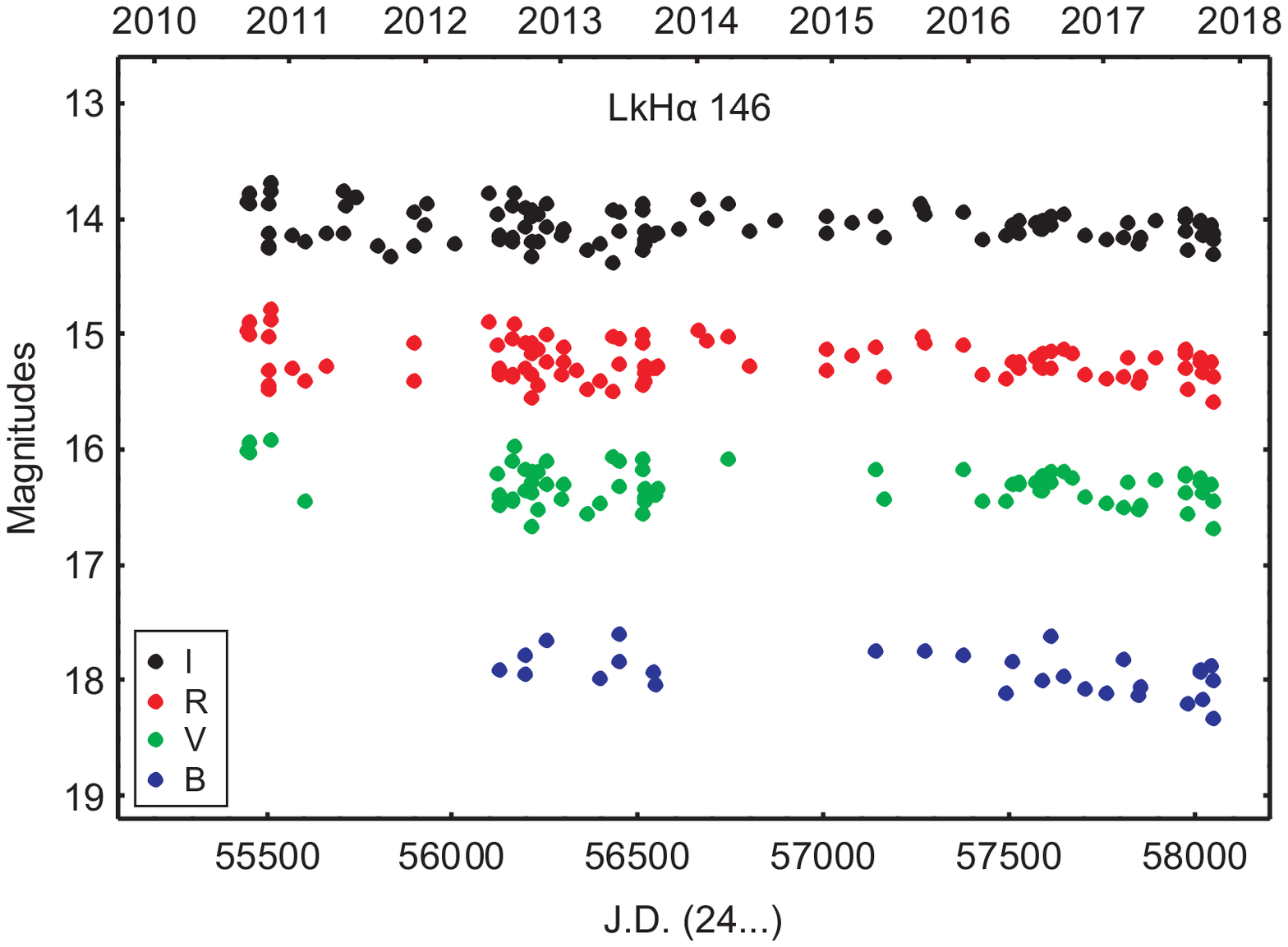}
\includegraphics[width=7cm]{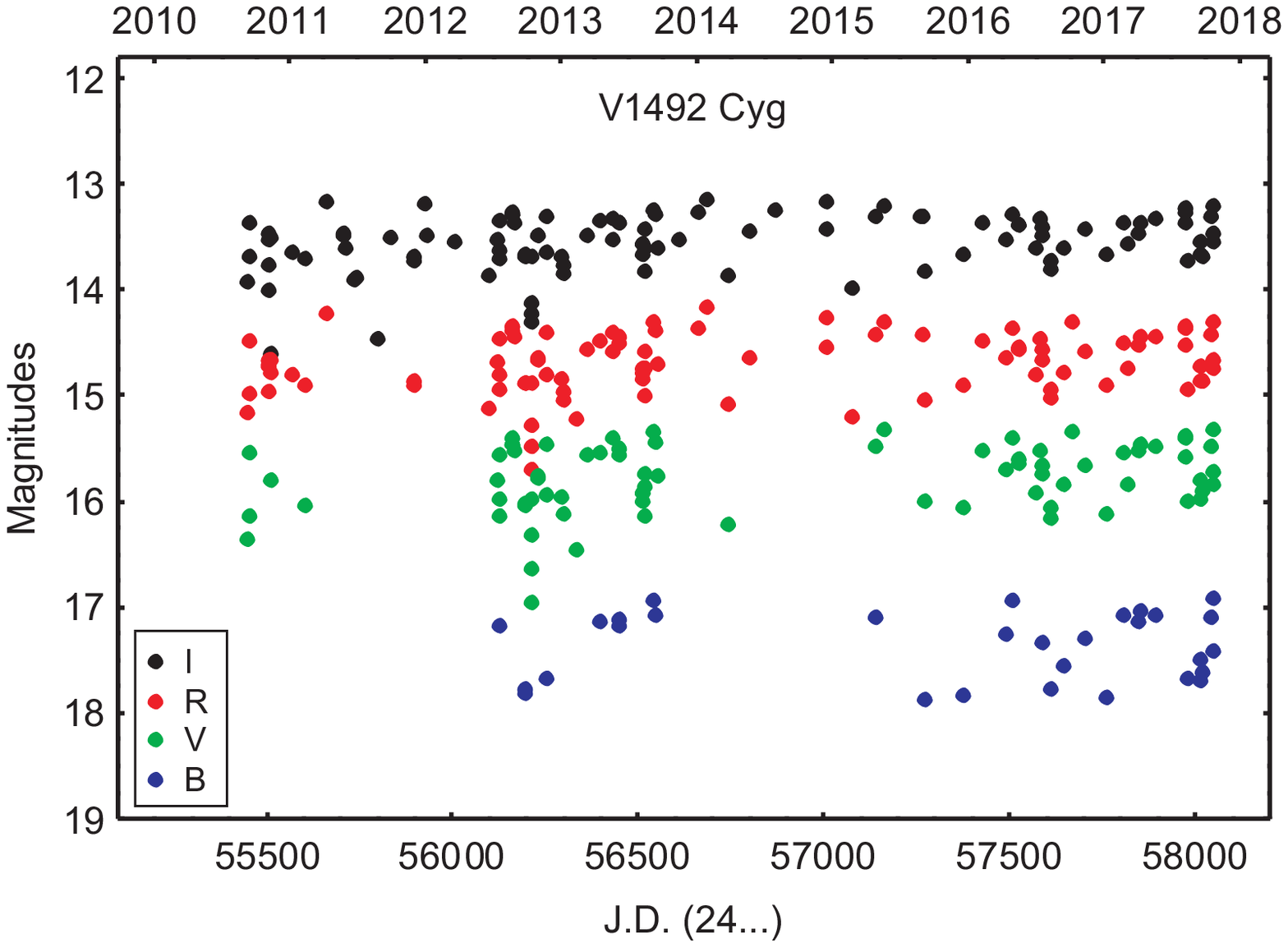}
\caption{Multicolor light curves of LkH$_{\alpha}$ 137, V1490 Cyg, LkH$_{\alpha}$ 146 and V1492 Cyg.}\label{Fig:subsec1}
\end{center}
\end{figure}

The spectrogram of LkH$_{\alpha}$ 137 taken by Herbig~(\cite{herb58}) shows bright H$_{\beta}$ and H$_{\gamma}$ on a continuous spectrum too weak for classification.
According to the author of direct photographs, the image of LkH$_{\alpha}$ 137 is diffuse, as if it were closely nebulous.
Ogura et al.~(\cite{ogur02}) detected H$_{\alpha}$ emission in the spectrum of V1490 Cyg.
The light curves of LkH$_{\alpha}$ 137, V1490 Cyg, LkH$_{\alpha}$ 146 and V1492 Cyg constructed on the basis of our observations are plotted on Fig.~\ref{Fig:subsec1}.

It can be seen from Fig.~\ref{Fig:subsec1} that LkH$_{\alpha}$ 137 spends most of the time at high light.
During our photometric monitoring, one decline in the star's brightness is registered.
The observed drop began in mid-2012 and it reached the biggest depth in 2013 April with amplitude 1.40 mag in $B$-band, 1.18 mag in $V$-band, 1.01 mag in $R$-band and 0.70 mag in $I$-band.
In 2013 May the star's brightness returned to the previous high level.

\newpage
The measured color indices ($V-I$, $V-R$ and $B-V$) versus the star's $V$ magnitude are plotted on Fig.~\ref{Fig:colors}.
It can be seen from Fig.~\ref{Fig:colors} that LkH$_{\alpha}$ 137 becomes redder as it fades.
Normally the star becomes redder when its light is being obscured by dust clumps or filaments in the line of sight.
On the 2MASS color$-$color diagram (Fig.~\ref{Fig:2mass}) LkH$_{\alpha}$ 137 lies close to the CTTSs location, i.e. the star has infrared excess indicating the presence of disk around it.
Therefore, the probable cause for the observed drop in the brightness of LkH$_{\alpha}$ 137 is the variation in the density of the dust in orbit around the star, which crosses the line of sight and obscures the object.
Evidence of periodicity in the brightness variability of LkH$_{\alpha}$ 137 is not detected.

\newpage
V1490 Cyg and V1492 Cyg also spend most of the time at high light (Fig.~\ref{Fig:subsec1}).
The stars exhibit irregular multiple fading events in all bands with different amplitudes.
The time periods of the declines in the brightness are relatively short, usually we have only one photometric point in some minima.
The registered amplitudes of the brightness variations of V1490 Cyg and V1492 Cyg during the whole time of observations are given in Table~\ref{Tab:amplitudes}.
On the 2MASS diagram (Fig.~\ref{Fig:2mass}) V1490 Cyg and V1492 Cyg are located close to the CTTSs line.
The measured color indices versus the stellar $V$ magnitude for the objects are plotted on Fig.~\ref{Fig:colors}.
It is seen from the figure that the stars become redder as they fade.
Such color variations are typical for T Tauri variables, whose variability is produced by small irregular obscuration by the circumstellar material or by the rotational modulation of one or more spots on the stellar surface.
Evidence of periodicity in the brightness variability of V1490 Cyg and V1492 Cyg is not detected.

\begin{figure}
   \centering
   \includegraphics[width=7cm, angle=0]{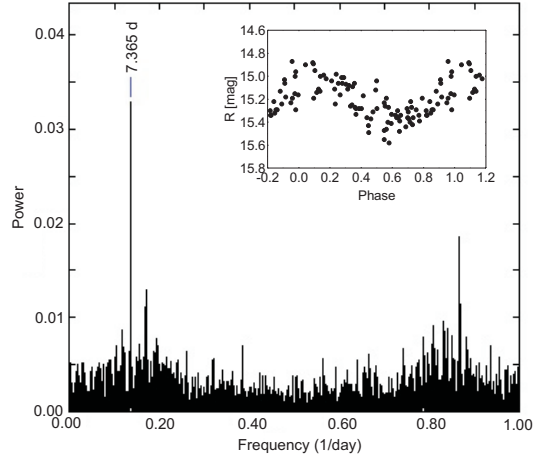}
   \caption{Periodogram of the photometric data of LkH$_{\alpha}$ 146.}\label{Fig:periodLkHa146}
   \end{figure}

LkH$_{\alpha}$ 146 shows both active states with high amplitude and quiet states with lower amplitude, at the same brightness level.
The registered amplitudes of the brightness variations of the star during our photometric monitoring are given in Table~\ref{Tab:amplitudes}.
An important result from our study of LkH$_{\alpha}$ 146 is the registered previously unknown periodicity in its photometric behavior.
On Fig.~\ref{Fig:periodLkHa146} we show the obtained periodogram using all data in $R$-band of the star.
We found a significant peak in the periodogram corresponding to 7.365 day period.
False Alarm Probability estimation was done by randomly deleting about 10\% of the data for about 20 times and then returning the period of determination.
The period remain stable during the whole time of our observations (2010$-$2017), even with a subsample with 20\% of the data removed.
The phase-folded light curve of LkH$_{\alpha}$ 146 in the $R$-band according to 7.365 day period is also plotted on Fig.~\ref{Fig:periodLkHa146}.

\newpage
The period we found is a typical rotational period of a young low-mass star.
The periodicity could be caused by rotation modulation of dark spots on the stellar surface.
According to Herbst et al. (1994) dark spots may last for hundreds or thousands of rotations of the star, as in the case of LkH$_{\alpha}$ 146.

\subsection{2MASS J20504608+4419100, V1597 Cyg, V1598 Cyg and V1956 Cyg}

In the literature these objects were reported as flare stars.
2MASS J20504608+4419100 was discovered and classified as a flare star by Jankovics et al.~(\cite{jank80}).
The authors reported a flare event on 1979 August 28 when the brightness of the star increased with an amplitude of 3.2 mag in $U$-light.
Erastova \& Tsvetkov~(\cite{eras74}) reported a flare event on 1973 August 24 in V1597 Cyg when the star's brightness increased with an amplitude of 3.0 mag in $U$-light.
Tsvetkov et al.~(\cite{tsve74}) registered a flare event on 1972 August 16 in the photometric behavior of V1598 Cyg with an amplitude of 1.7 in $U$-light.
Rosino et al.~(\cite{rosi87}) also reported a flare event in V1598 Cyg on 1972 October 5 with an amplitude of 2.6 in $U$-light.
K\'{o}sp\'{a}l et al.~(\cite{kosp11}) used V1598 Cyg as one of the comparison stars in their study of VSX J205126.1+440523 (V2492 Cyg).
V1956 Cyg was classified as a flare star by Tsvetkov \& Tsvetkova~(\cite{tsve90}).

During our photometric monitoring we did not register flare events in these stars.
Our negative result does not mean that these objects do not show flares.
Our mode of observations (about 100 estimations for $\sim$2550 nights) does not allow to find out any flares.
For detection of flares, it is necessary to undertake continuous monitoring during several tens of nights with the time resolution in some seconds.

\begin{figure}
\begin{center}
\includegraphics[width=7cm]{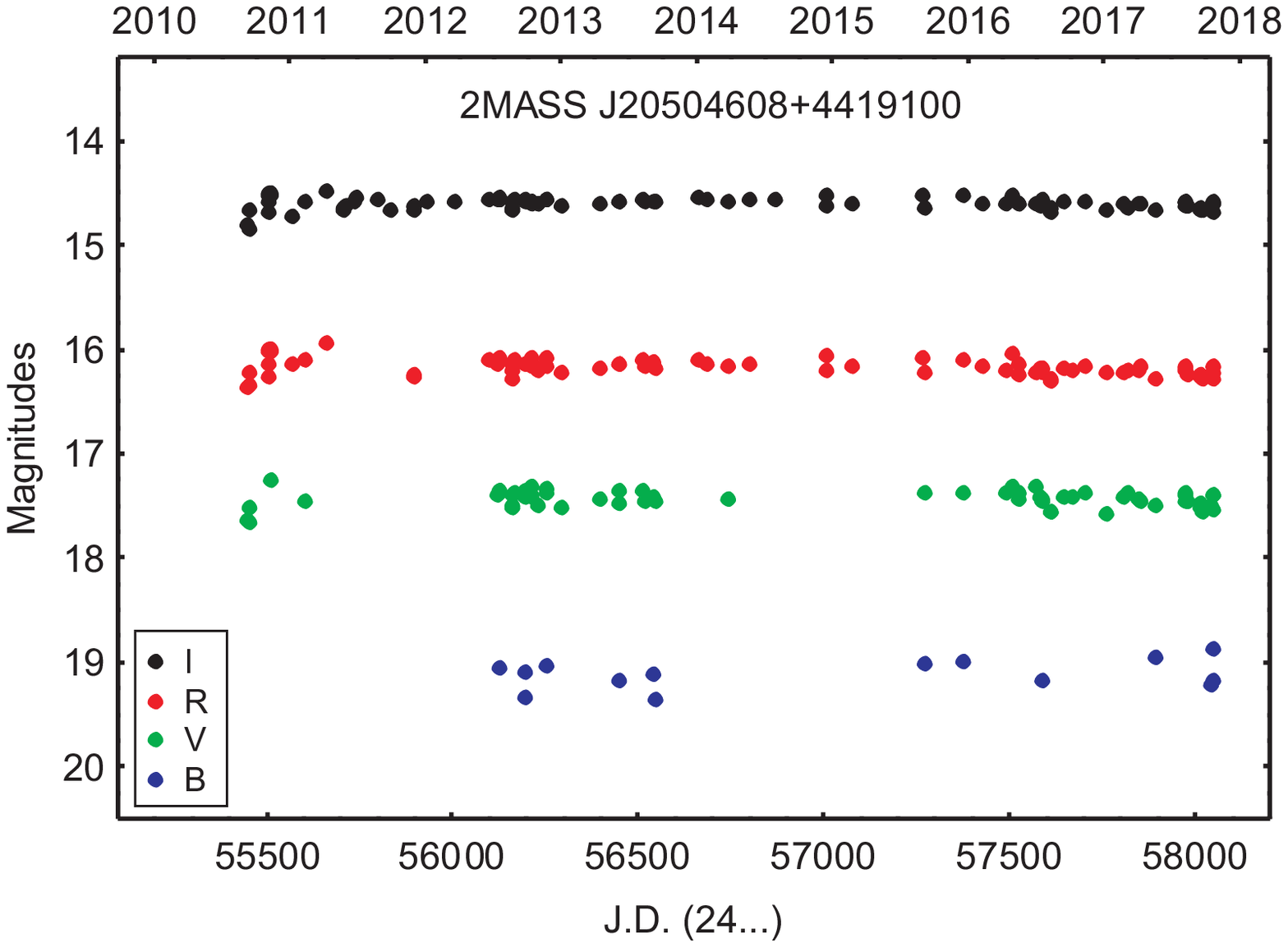}
\includegraphics[width=7cm]{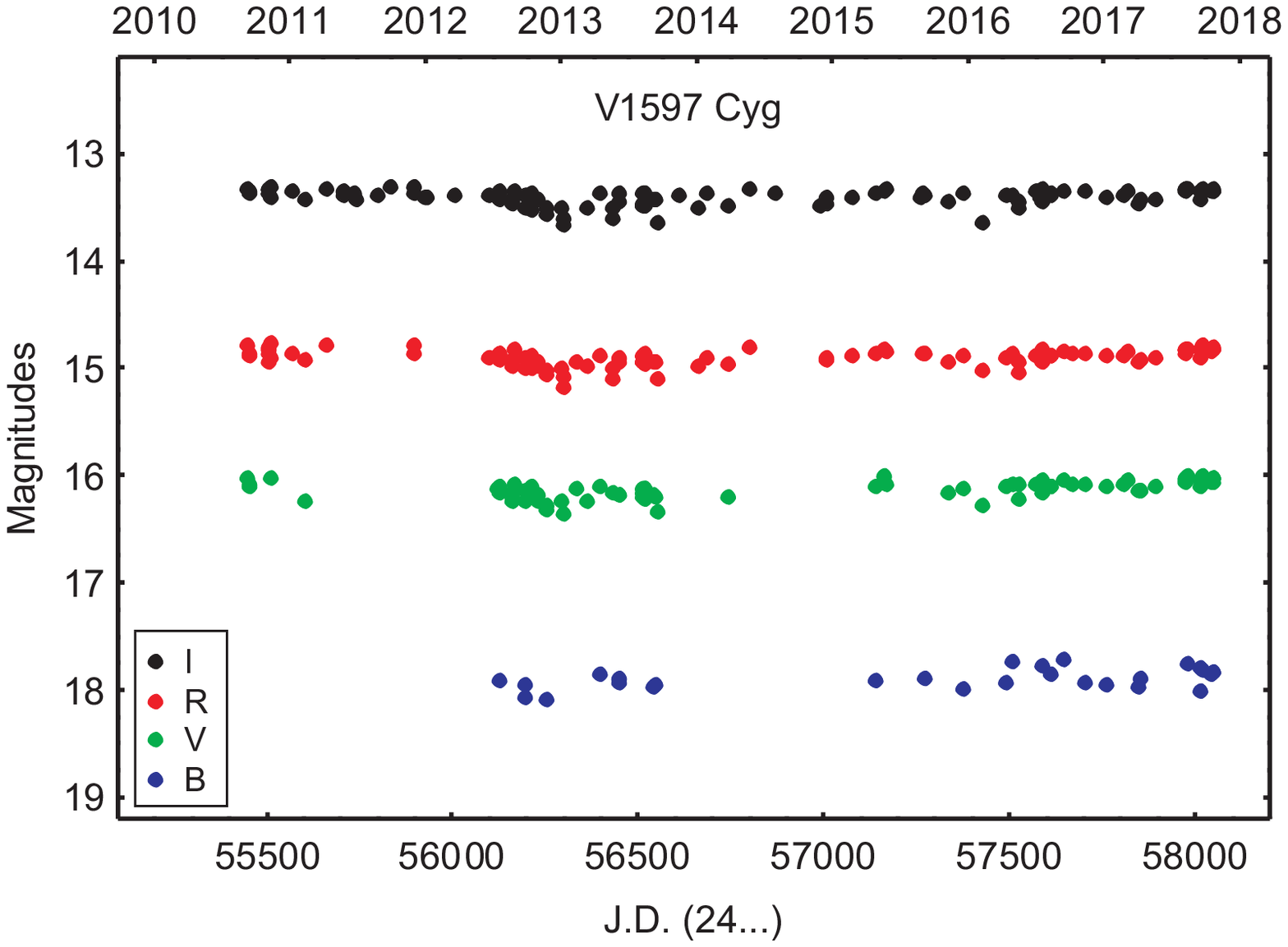}
\includegraphics[width=7cm]{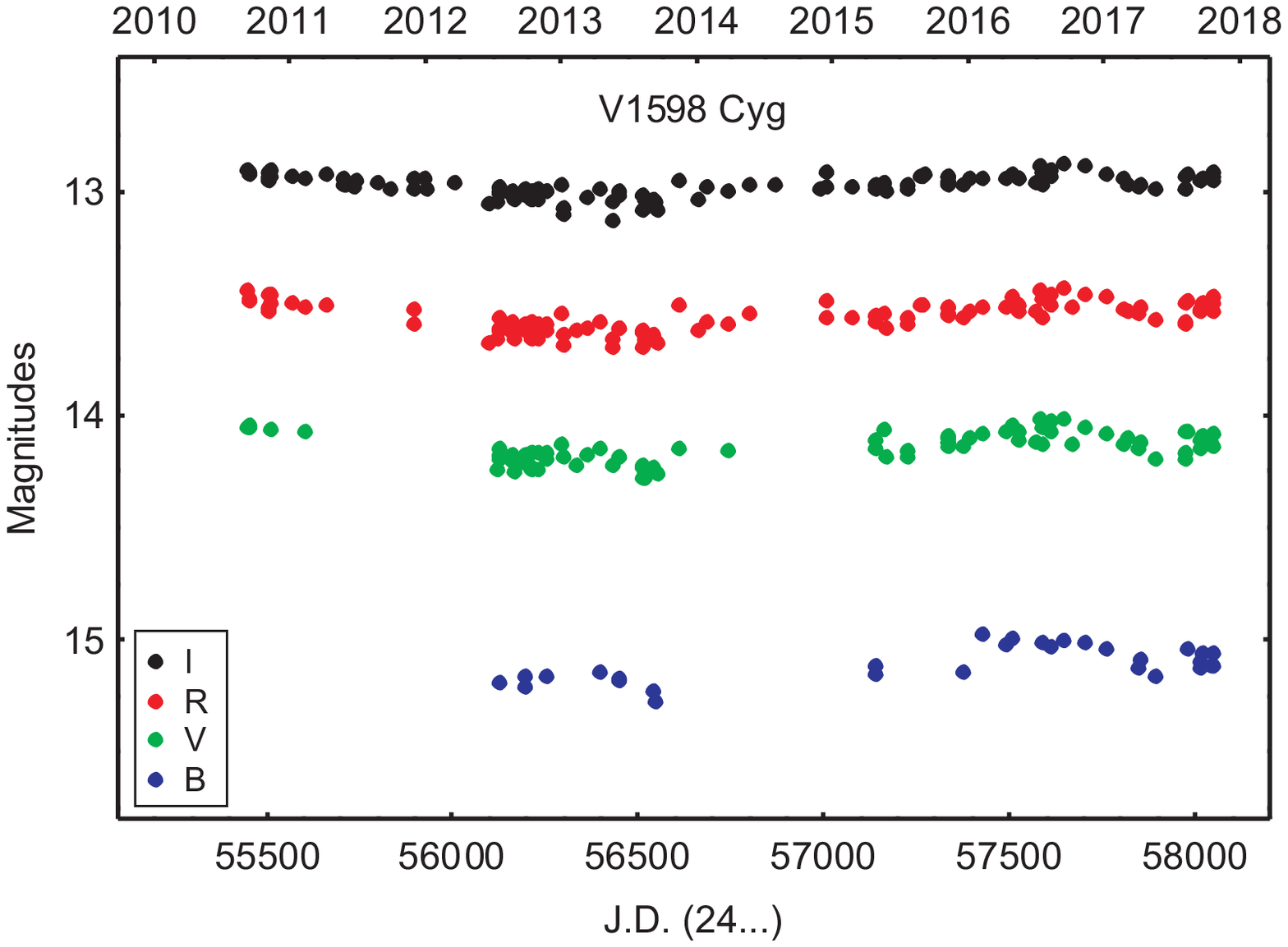}
\includegraphics[width=7cm]{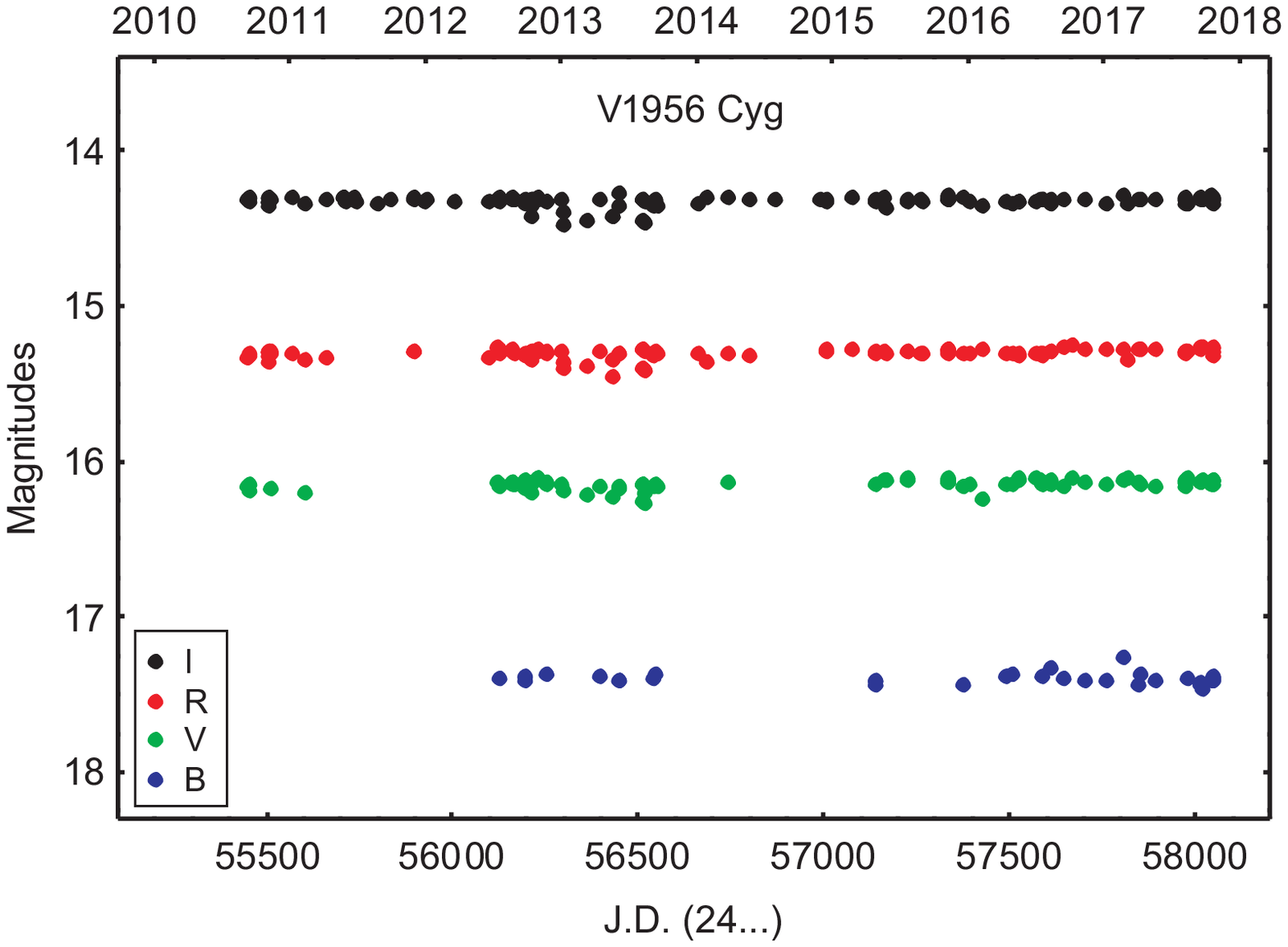}
\caption{Multicolor light curves of 2MASS J20504608+4419100, V1597 Cyg, V1598 Cyg and V1956 Cyg.}\label{Fig:subsec2}
\end{center}
\end{figure}

$BVRI$ light curves of 2MASS J20504608+4419100, V1597 Cyg, V1598 Cyg and V1956 Cyg are plotted on Fig.~\ref{Fig:subsec2}.
The registered amplitudes of variability of the stars are given in Table~\ref{Tab:amplitudes}.
It can be seen that these objects show no significant photometric variability.
On the 2MASS diagram (Fig.~\ref{Fig:2mass}) the objects are located in different areas $-$ 2MASS J20504608+4419100 and V1597 Cyg lie above the CTTSs line while V1598 Cyg and V1956 Cyg lie close to both the main sequence line and the giant stars line.

\newpage
It is possible V1598 Cyg and V1956 Cyg to be projecting stars on the front of the field of NGC 7000/IC 5070 complex.
Using the data from our photometric observations it is very difficult to classify these stars.
Evidence of periodicity in their brightness variability is not detected.

\subsection{LkH$_{\alpha}$ 141, V1532 Cyg, LkH$_{\alpha}$ 161, LkH$_{\alpha}$ 172 and LkH$_{\alpha}$ 173} 

These objects are included in the list of H$_{\alpha}$ emission-line stars published by Herbig~(\cite{herb58}).
Guieu et al.~(\cite{guie09}) included these stars in their list of YSO candidates.
The spectrogram of LkH$_{\alpha}$ 172 taken by Herbig~(\cite{herb58}) shows an absorption spectrum of intermediate or late G type, with rather a strong emission at H$_{\beta}$.
According to the author emission may also be present at H$_{\gamma}$.
Terranegra et al.~(\cite{terr94}) determined the spectral type of LkH$_{\alpha}$ 172 as G3 III.

\begin{figure}
\begin{center}
\includegraphics[width=7cm]{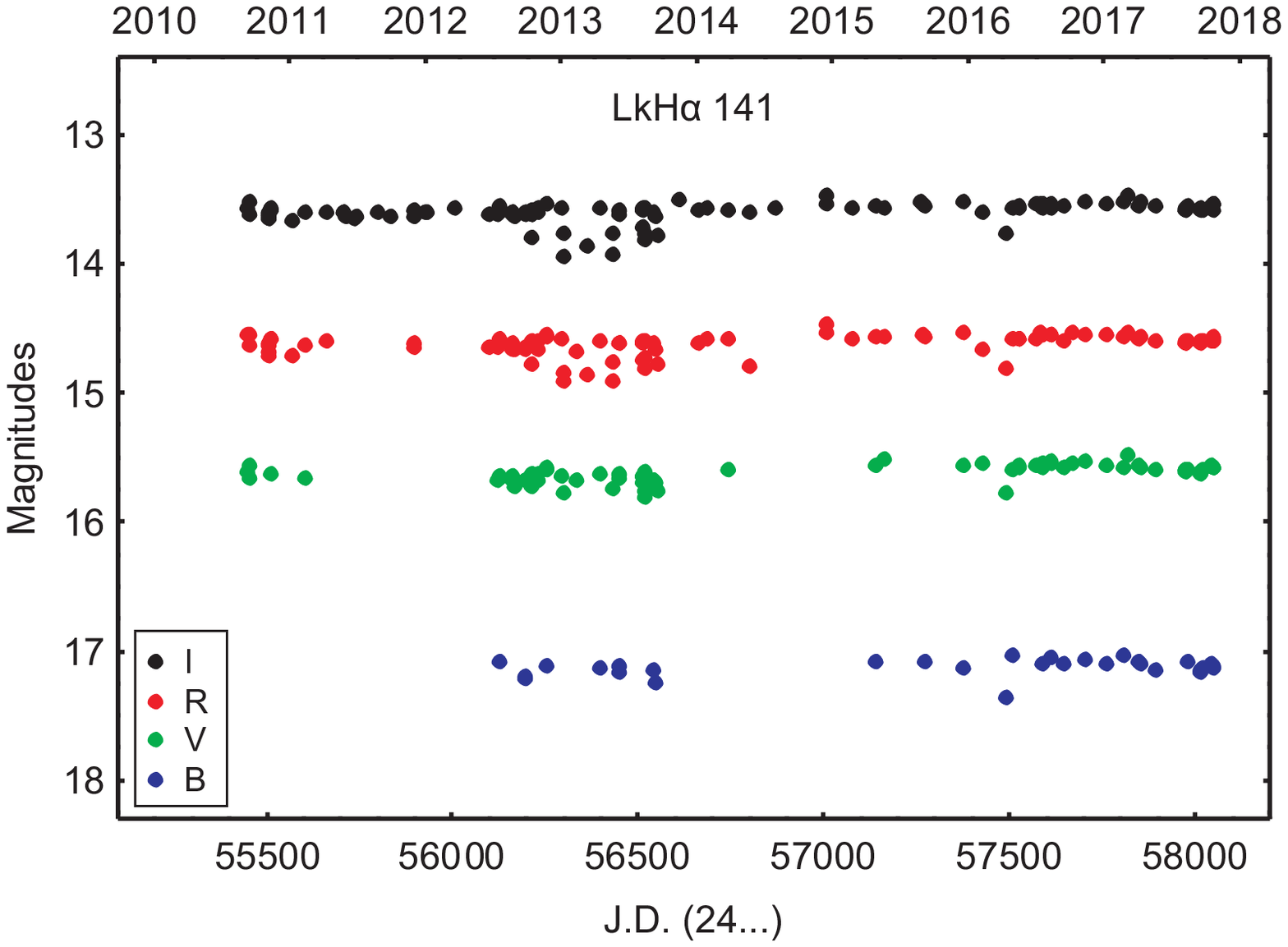}
\includegraphics[width=7cm]{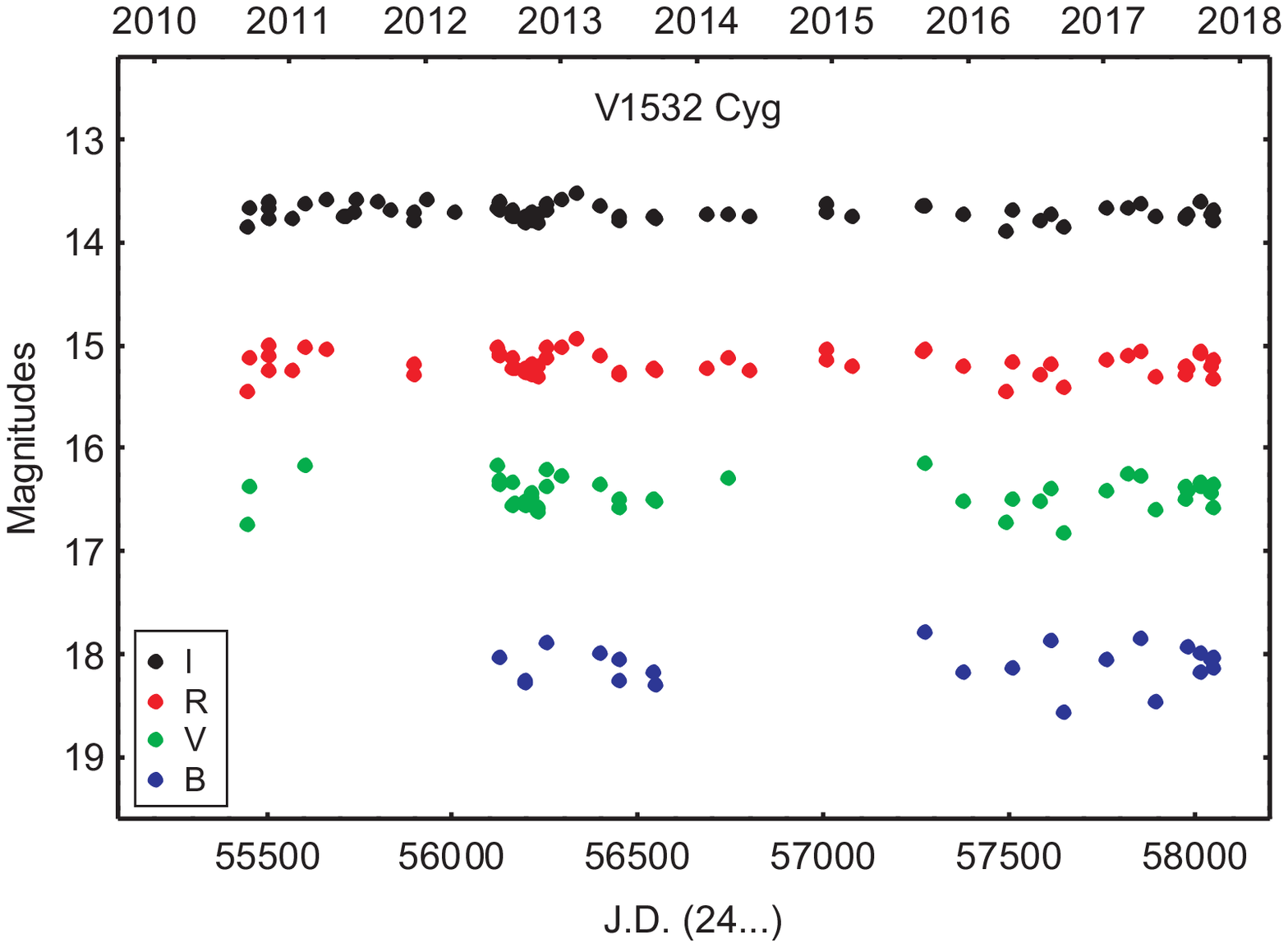}
\includegraphics[width=7cm]{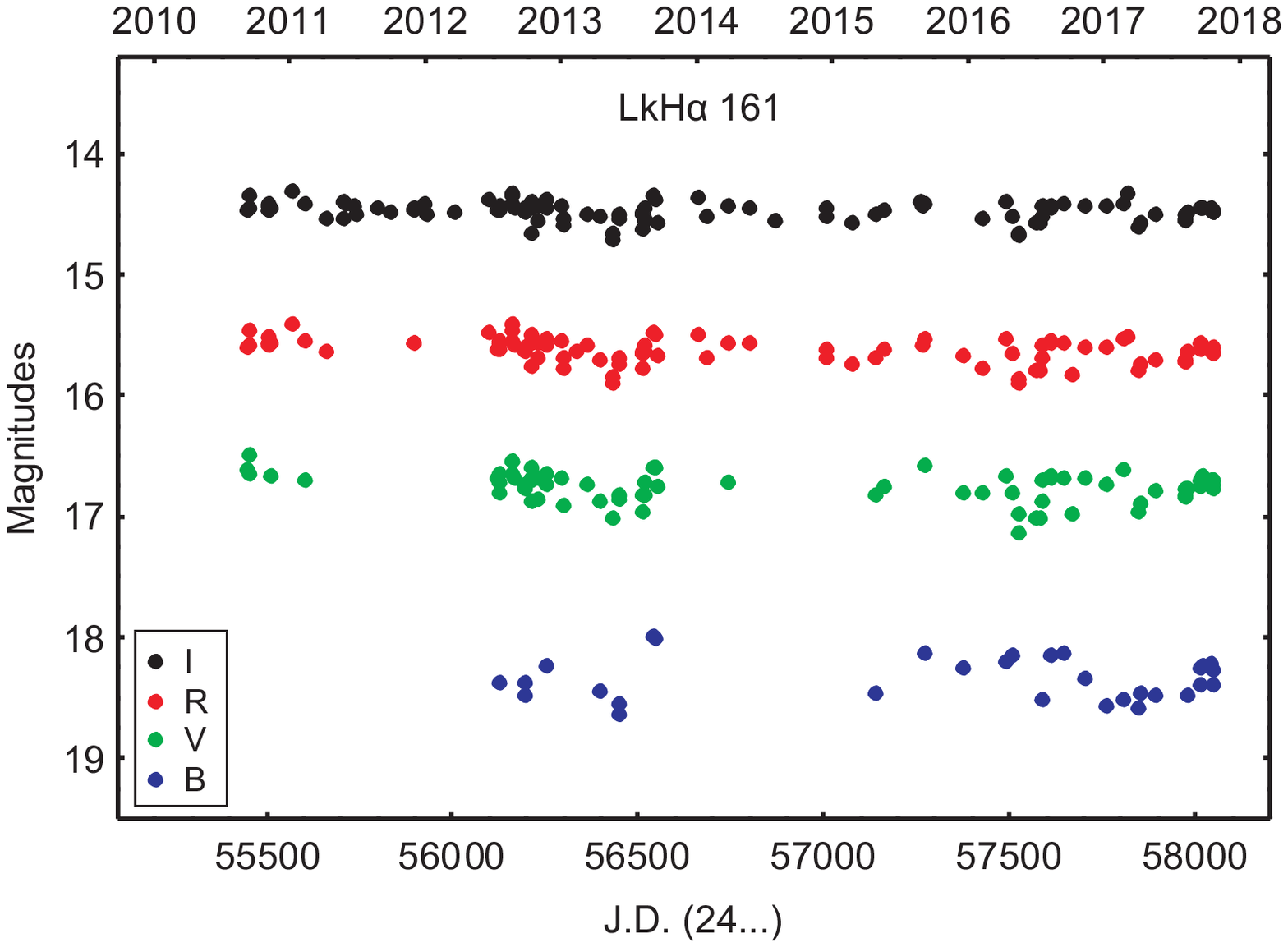}
\includegraphics[width=7cm]{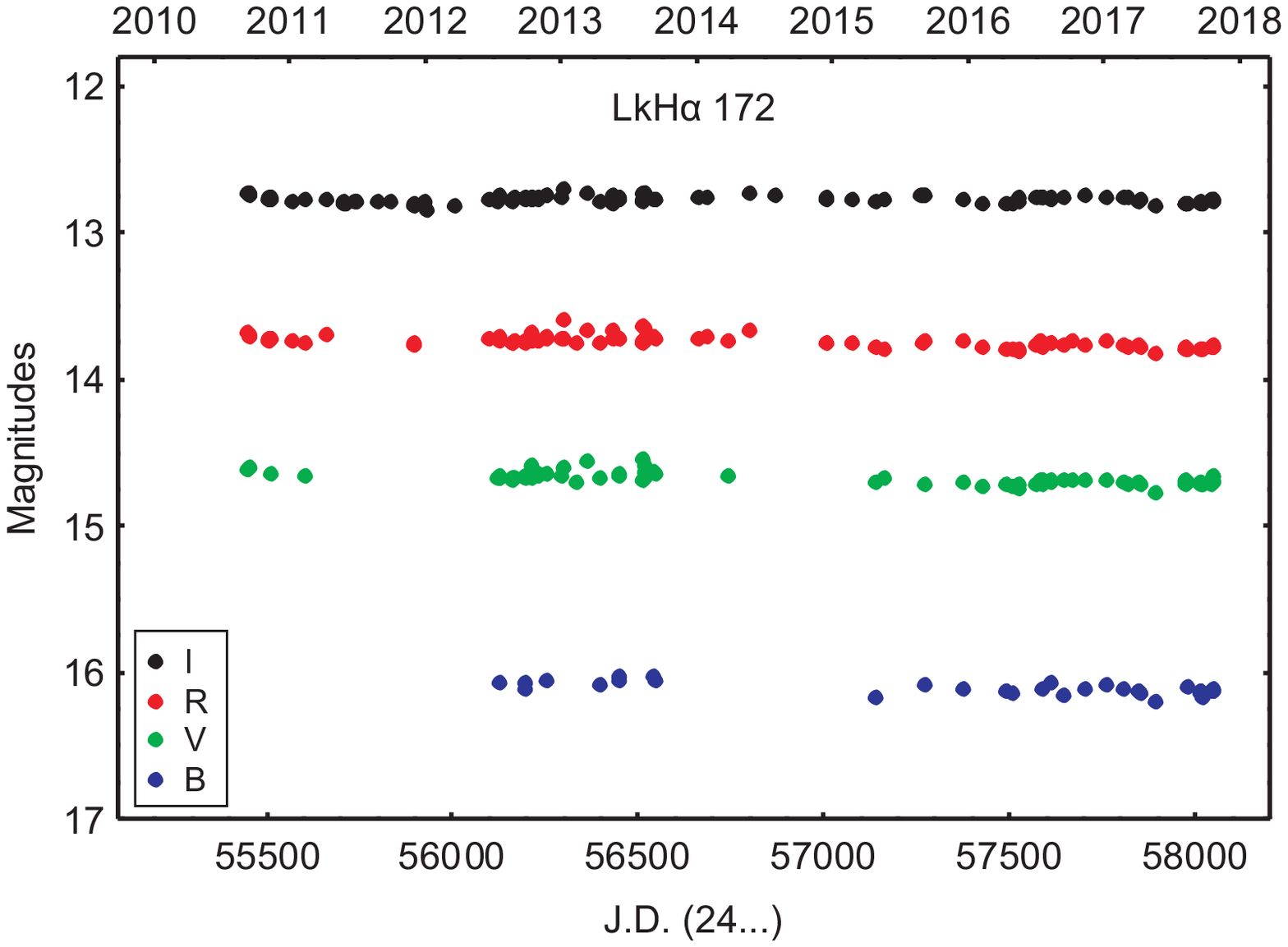}
\includegraphics[width=7cm]{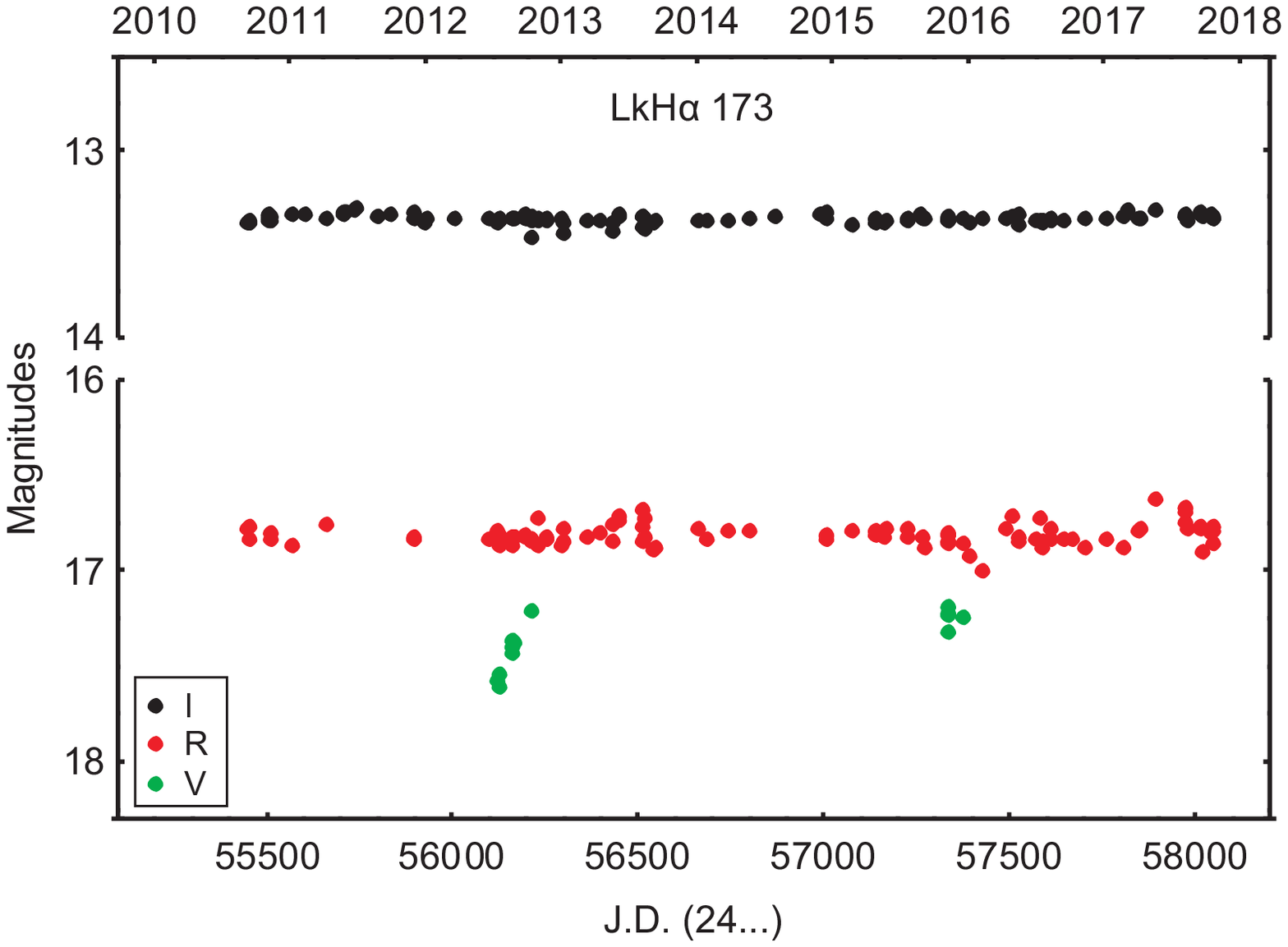}
\caption{Multicolor light curves of LkH$_{\alpha}$ 141, V1532 Cyg, LkH$_{\alpha}$ 161, LkH$_{\alpha}$ 172 and LkH$_{\alpha}$ 173.}\label{Fig:subsec3}
\end{center}
\end{figure}

The results from our photometric monitoring indicate that LkH$_{\alpha}$ 141, V1532 Cyg, LkH$_{\alpha}$ 161, LkH$_{\alpha}$ 172 and LkH$_{\alpha}$ 173 show photometric variability with small amplitudes.
The registered amplitudes of variability of the stars from the current section are given in Table~\ref{Tab:amplitudes}.
$BVRI$ light curves of LkH$_{\alpha}$ 141, V1532 Cyg, LkH$_{\alpha}$ 161, LkH$_{\alpha}$ 172 and LkH$_{\alpha}$ 173 are plotted on Fig.~\ref{Fig:subsec3}.
On the 2MASS color$-$color diagram (Fig.~\ref{Fig:2mass}) the objects are located close to the CTTSs line.
These stars have infrared excess indicating the presence of circumstellar disks around them.
Probably the brightness variability of the objects is caused by the variations in the mass accretion rate and/or the modulation of the stellar brightness in the presence of spot(s) on the stellar surface.
The registered amplitudes of photometric variability also confirmed this suspicion (see Herbst et al.~\cite{herb94}).
Evidence of periodicity in the brightness variability of the stars is not detected.

\newpage
\subsection{LkH$_{\alpha}$ 147 and LkH$_{\alpha}$ 168}

In the literature LkH$_{\alpha}$ 147 and LkH$_{\alpha}$ 168 are classified as HAEBESs.
The spectrogram of LkH$_{\alpha}$ 147 taken in 1957 by Herbig~(\cite{herb58}) shows an essentially continuous spectrum with emission at H$_{\beta}$.
According to the author although no definite spectral type can be assigned, there is no doubt that this is an early-type star.
Hern\'{a}ndez et al.~(\cite{hern04}) classified LkH$_{\alpha}$ 147 as a HAEBES and estimated its spectral type as B2.
Herbst \& Shevchenko~(\cite{herb99}) included the star in their list of HAEBES and measured $\overline{V}$=14.45 mag, $\overline{B-V}$=1.50 mag and $\overline{V-R}$=1.61 mag of the object during the time period 2450007$-$2450017.

\newpage
According to Herbig~(\cite{herb58}) the spectral type of LkH$_{\alpha}$ 168 is intermediate or late B, with no emission apparent in the photographic region on a 1954 plate.
Chavarr\'{\i}a-K. et al.~(\cite{chav89}) concluded the spectral type of the star is F8-G0.
According to the authors LkH$_{\alpha}$ 168 has a warm circumstellar dust envelope of $\sim$1000 K and the star shows signs of photometric variability.
Shevchenko et al.~(\cite{shev93}) defined the spectral type of LkH$_{\alpha}$ 168 as A8-F2e.
Herbst \& Shevchenko~(\cite{herb99}) included the star in their list of HAEBES and measured $\overline{V}$=13.48 mag, $\overline{U-B}$=1.05 mag, $\overline{B-V}$=1.25 mag, and $\overline{V-R}$=1.21 mag of the object during the time period 2446255$-$2450017.

\begin{figure}
\begin{center}
\includegraphics[width=7cm]{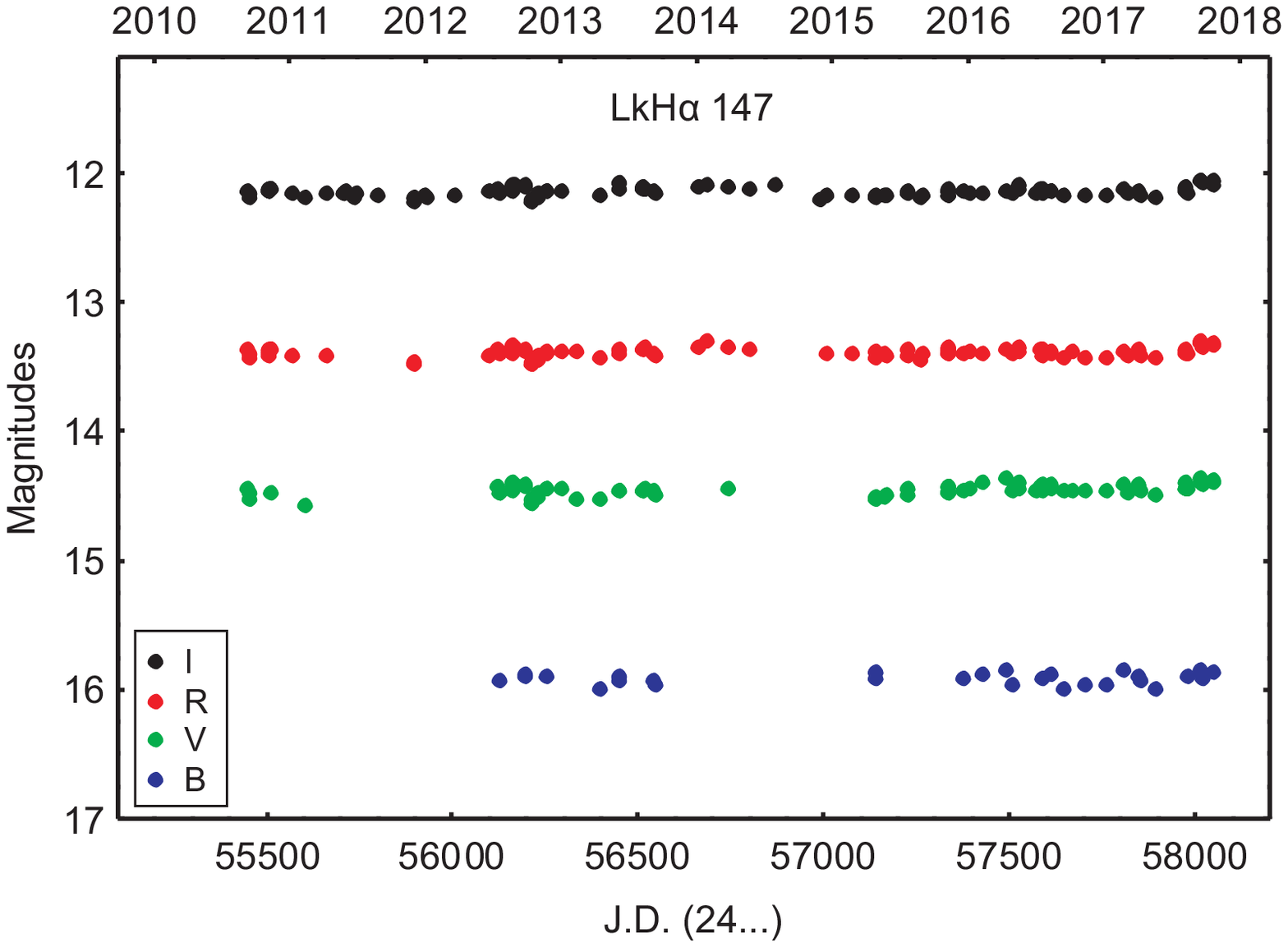}
\includegraphics[width=7cm]{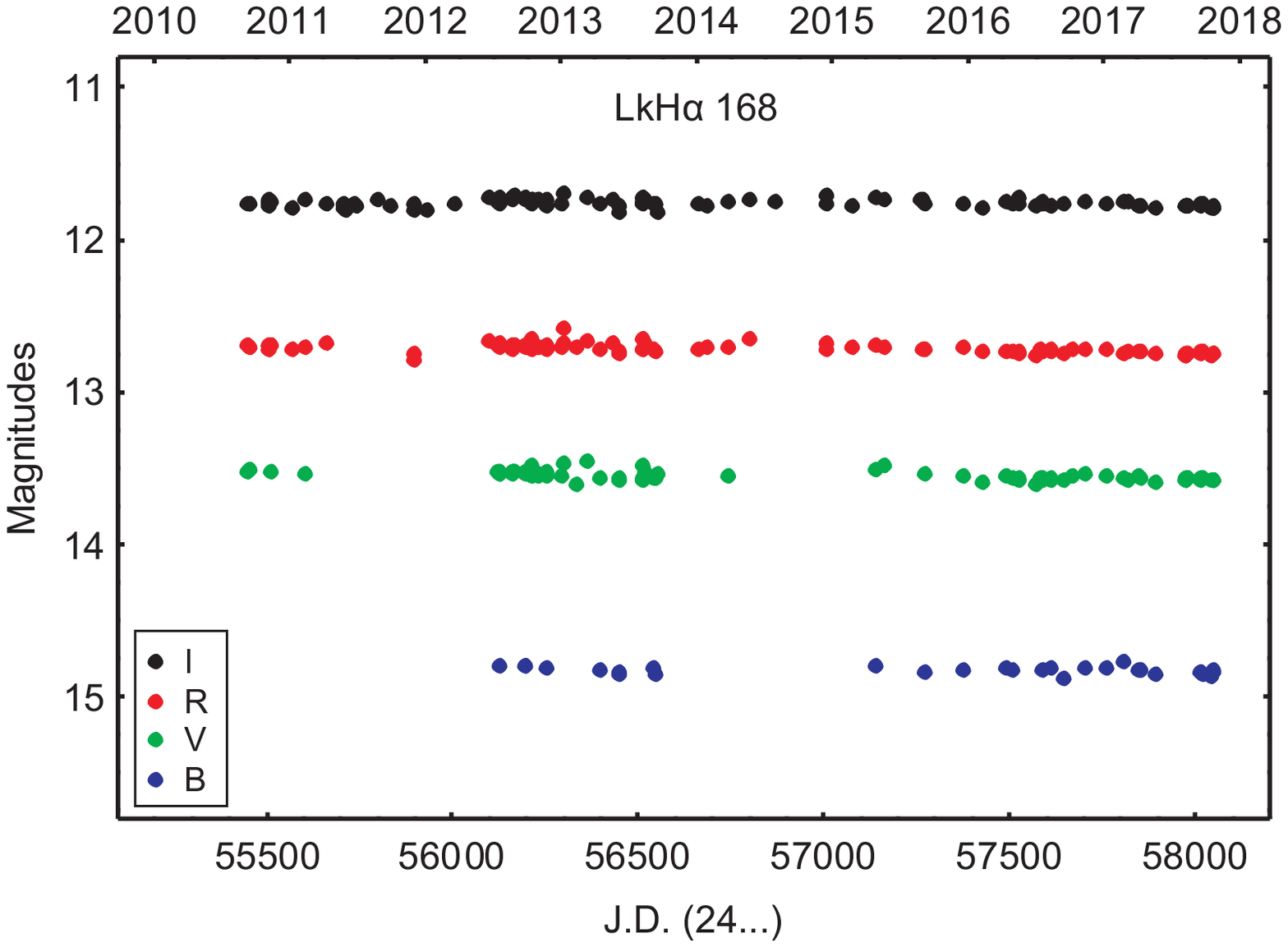}
\caption{Multicolor light curves of LkH$_{\alpha}$ 147 and LkH$_{\alpha}$ 168.}\label{Fig:subsec4}
\end{center}
\end{figure}

$BVRI$ light curves of LkH$_{\alpha}$ 147 and LkH$_{\alpha}$ 168 are plotted on Fig.~\ref{Fig:subsec4}.
During our photometric monitoring these objects show no significant photometric variability.
We registered $\overline{V}$=14.44 mag, $\overline{B-V}$=1.46 mag, $\overline{V-R}$=1.06, and $\overline{V-I}$=2.31 mag for LkH$_{\alpha}$ 147, and $\overline{V}$=13.54 mag, $\overline{B-V}$=1.28 mag, $\overline{V-R}$=0.84, and $\overline{V-I}$=1.79 mag for LkH$_{\alpha}$ 168.

Our estimations of a light and color indexes for LkH$_{\alpha}$ 147 and LkH$_{\alpha}$ 168 are very close to estimations from work of Herbst \& Shevchenko~(\cite{herb99}).
There is only a significant difference in $V-R$ values for the two stars.
Probably the reason for this distinction that the $V-R$ color index is in the Johnson system in work of Herbst \& Shevchenko~(\cite{herb99}), and it is in the Cousins system in the present paper.
We used the transformation described in Fernie~(\cite{fern83}) and we obtained for $V-R$ color index from the work of Herbst \& Shevchenko~(\cite{herb99}) the value 1.06 mag for LkH$_{\alpha}$ 147 and 0.85 mag for LkH$_{\alpha}$ 168.
Apparently, the obtained values for $V-R$ color index after the transformation coincide with ones in present paper.

On the 2MASS diagram (Fig.~\ref{Fig:2mass}) LkH$_{\alpha}$ 147 is located close to the main sequence line while LkH$_{\alpha}$ 168 lies on the CTTSs line.
Evidence of periodicity in the brightness variability of these stars is not detected.

\newpage
\section{Concluding remarks}
\label{sect:conclusion remarks}
We reported for the $BVRI$ photometric behavior of 15 PMS stars around V2492 Cyg in the IC 5070 star-forming region.
Our observations cover 7-year time span (2010$-$2017) and represent the first long-term photometric monitoring of the investigated objects.
The obtained results can be summarized as follows:
(i) LkH$_{\alpha}$ 137, V1490 Cyg, LkH$_{\alpha}$ 146 and V1492 Cyg show bigger amplitudes of photometric variability among the objects from our study.
The results indicate that probably these stars are CTTSs;
(ii) although 2MASS J20504608+4419100, V1597 Cyg, V1598 Cyg and V1956 Cyg were classified as flare stars in the literature, we did not register any such events.
Our negative result does not mean that these objects do not show flares;
(iii) during our photometric monitoring LkH$_{\alpha}$ 141, V1532 Cyg, LkH$_{\alpha}$ 147, LkH$_{\alpha}$ 161, LkH$_{\alpha}$ 168, LkH$_{\alpha}$ 172 and LkH$_{\alpha}$ 173 show no significant photometric variability;
(iv) for LkH$_{\alpha}$ 146 we identified a 7.365 d rotational periodicity in its photometric behavior.

Further both photometric and spectral observations would offer clearer insight into the physical nature of the objects from our study.

\begin{acknowledgements}
This research has made use of the NASA's Astrophysics Data System.
This publication makes use of data products from the Two Micron All Sky Survey, which is a joint project of the University of Massachusetts and the Infrared Processing and Analysis Center/California Institute of Technology, funded by the National Aeronautics and Space Administration and the National Science Foundation.
This study has made use of the SIMBAD database (Wenger et al.~\cite{weng00}) and the VizieR catalogue access tool (Ochsenbein et al.~\cite{ochs00}), operated at CDS, Strasbourg, France.
This work was partly supported by the National Science Fund of the Ministry of Education and Science of Bulgaria under grants DM 08-2/2016, DN 08-1/2016, DN 08-20/2016 and DN 18-13/2017 and by funds of the project RD-08-112/2018 of the University of Shumen.
\end{acknowledgements}

\appendix                  

\section{Color indices ($V-I$, $V-R$ and $B-V$) versus the stellar $V$-magnitude of the stars from our study}
\begin{figure}
\begin{center}
\includegraphics[width=4.5cm]{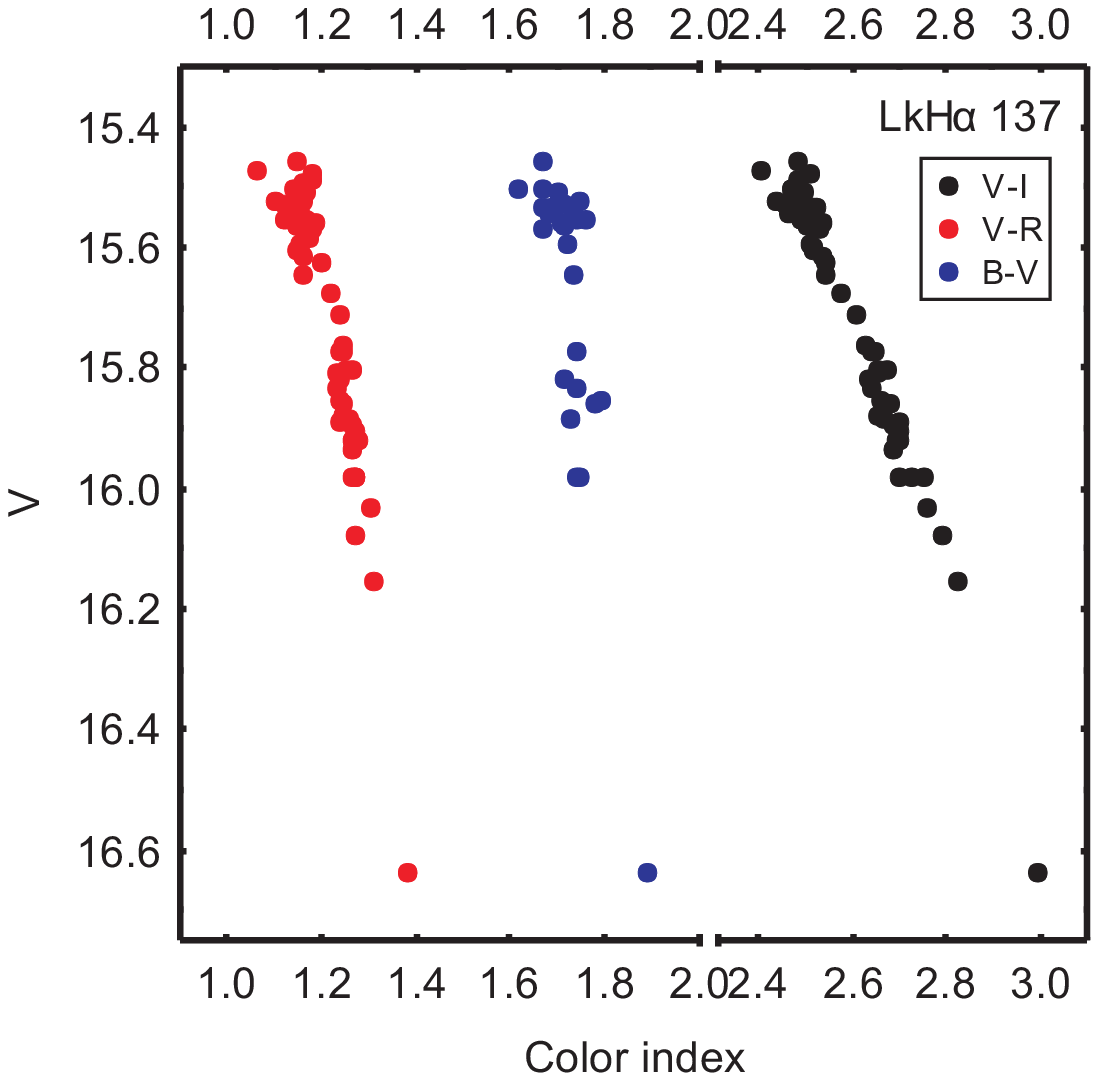}
\includegraphics[width=4.5cm]{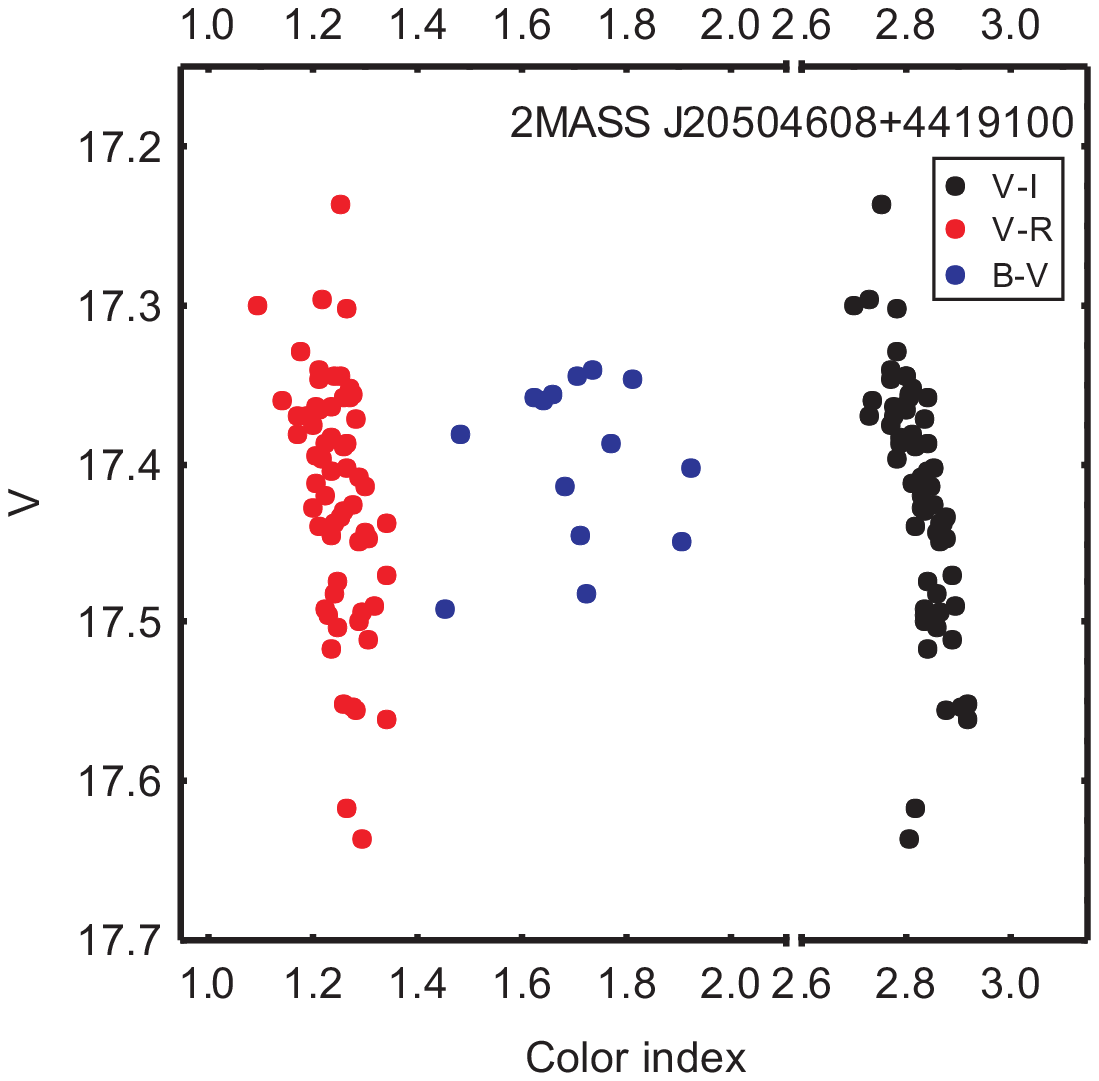}
\includegraphics[width=4.5cm]{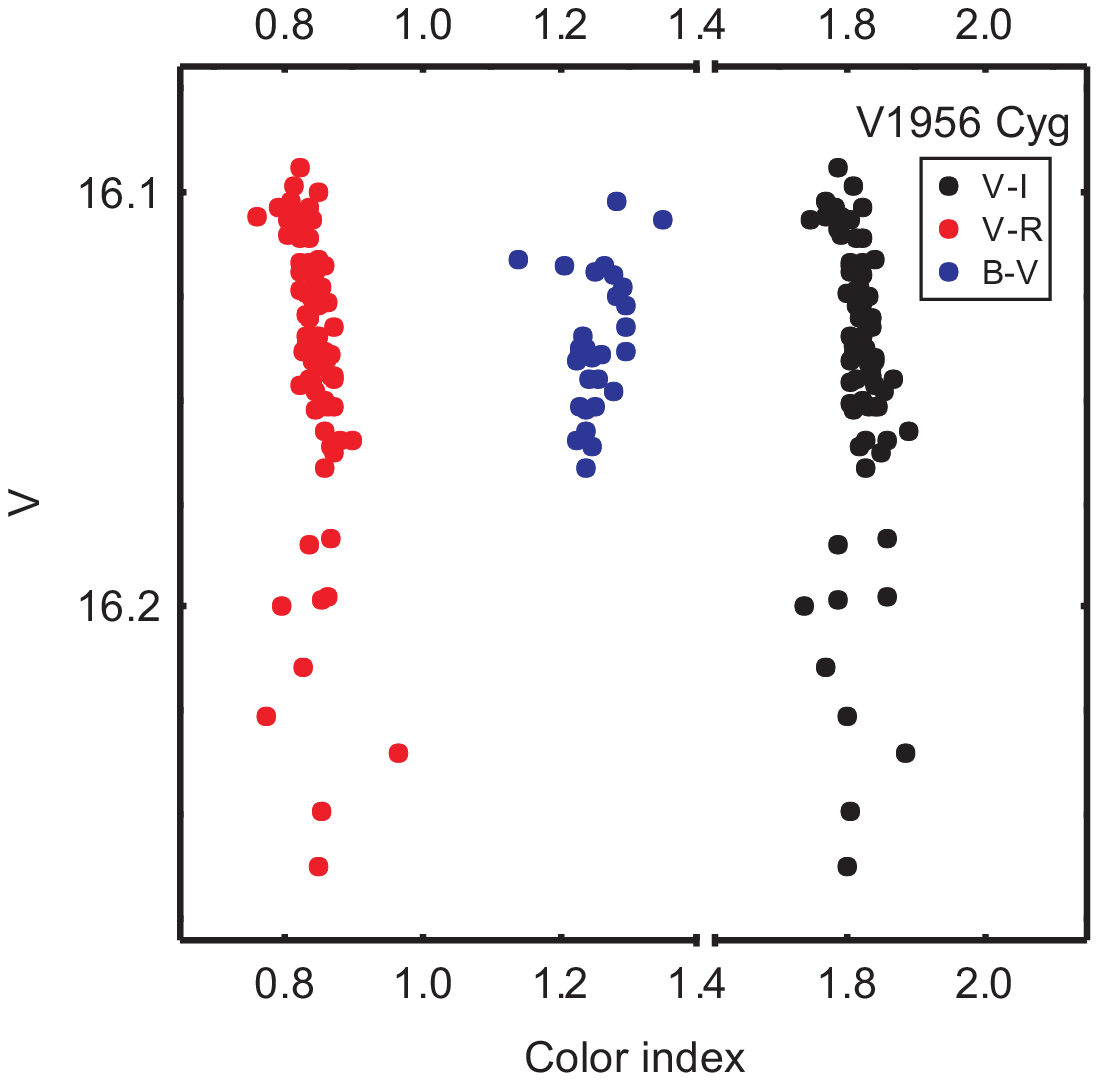}
\includegraphics[width=4.5cm]{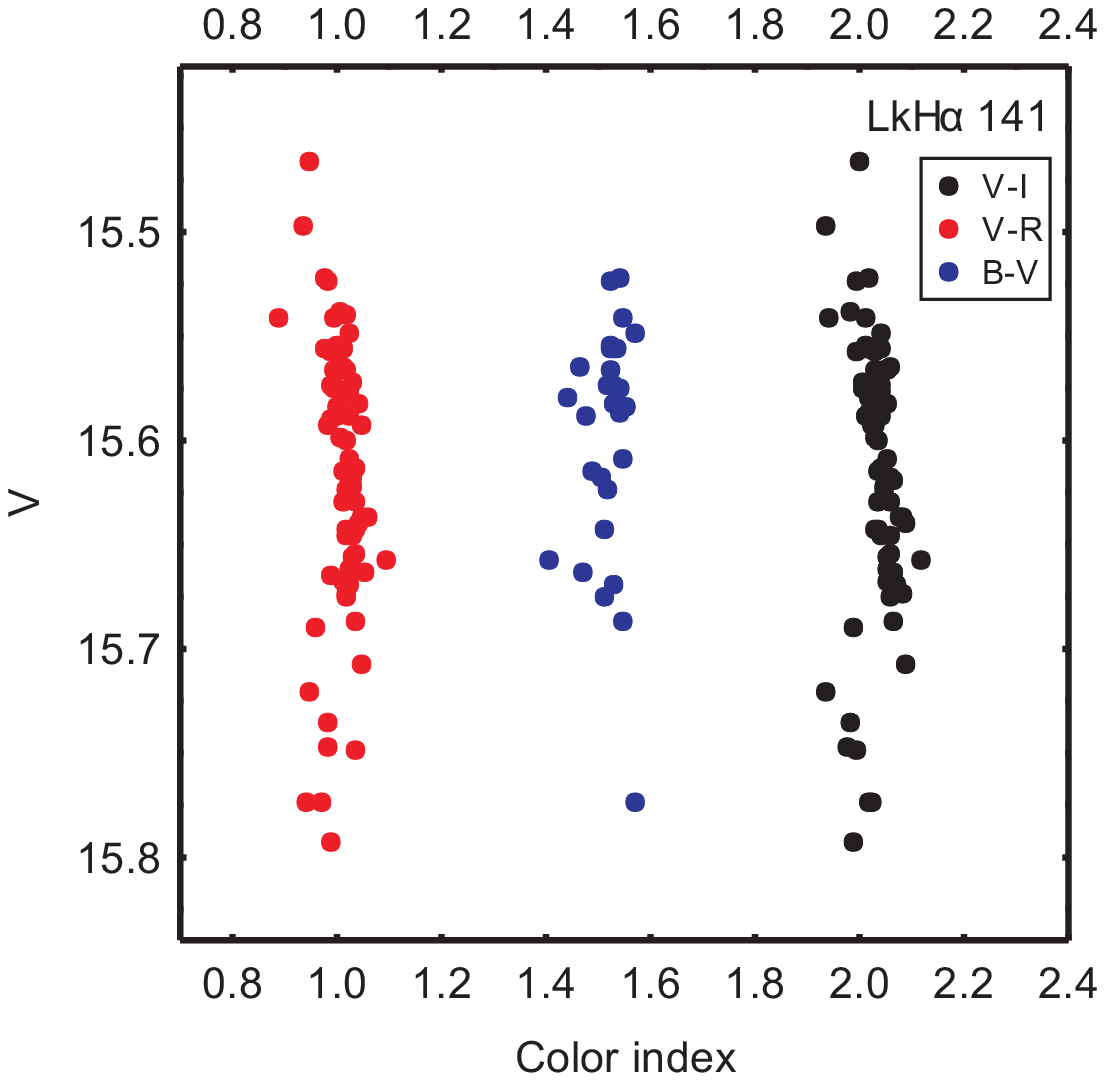}
\includegraphics[width=4.5cm]{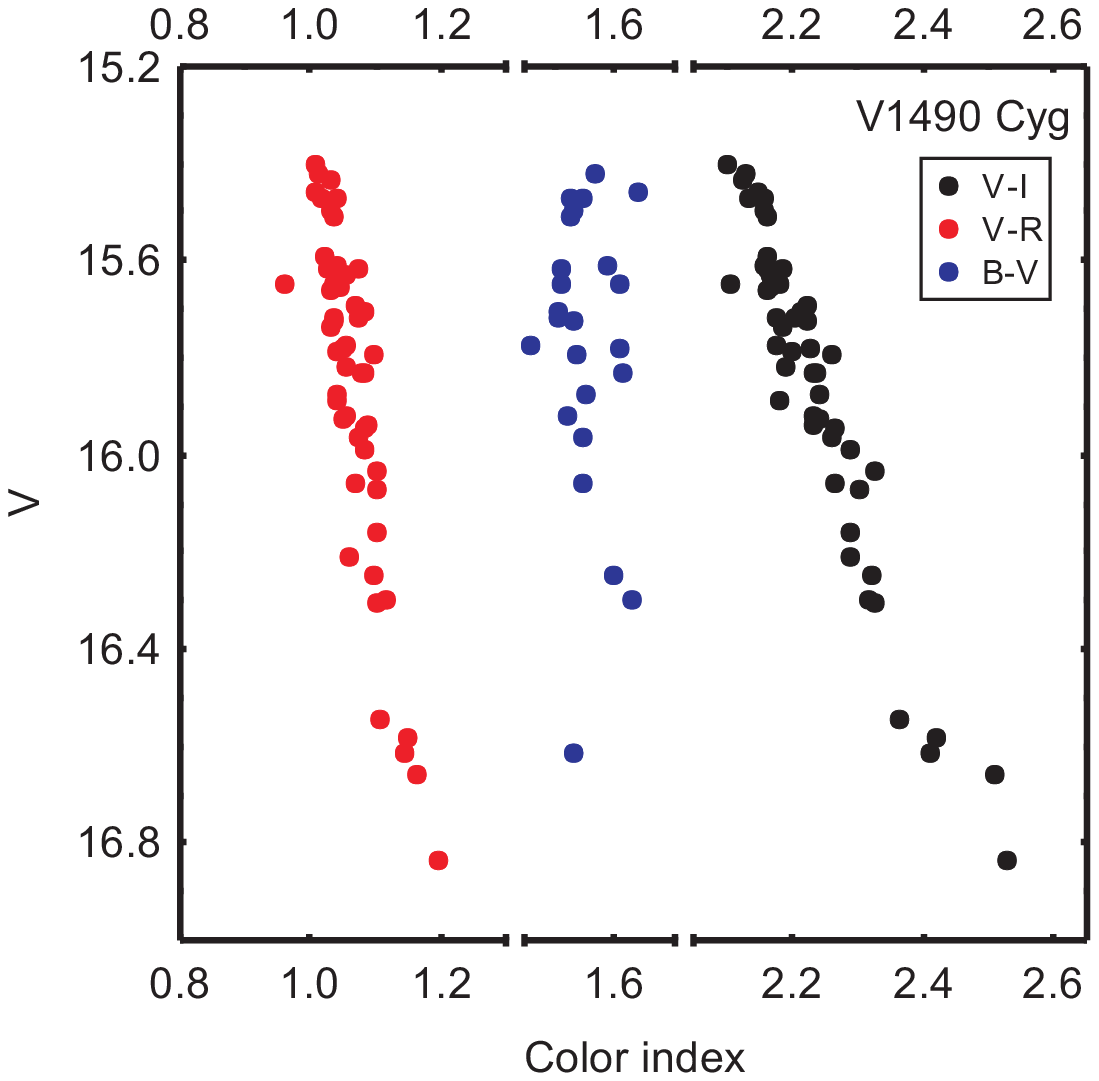}
\includegraphics[width=4.5cm]{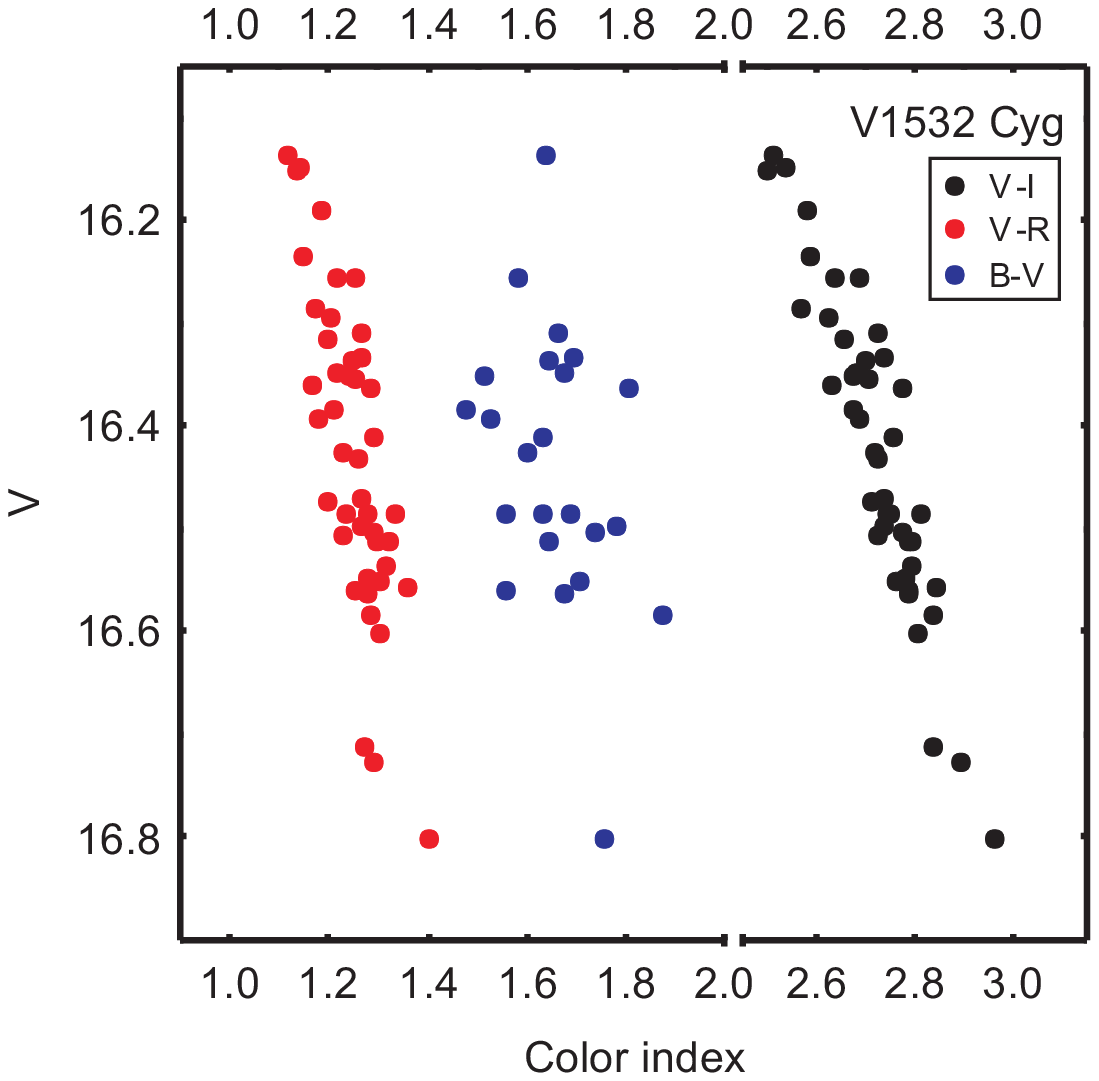}
\includegraphics[width=4.5cm]{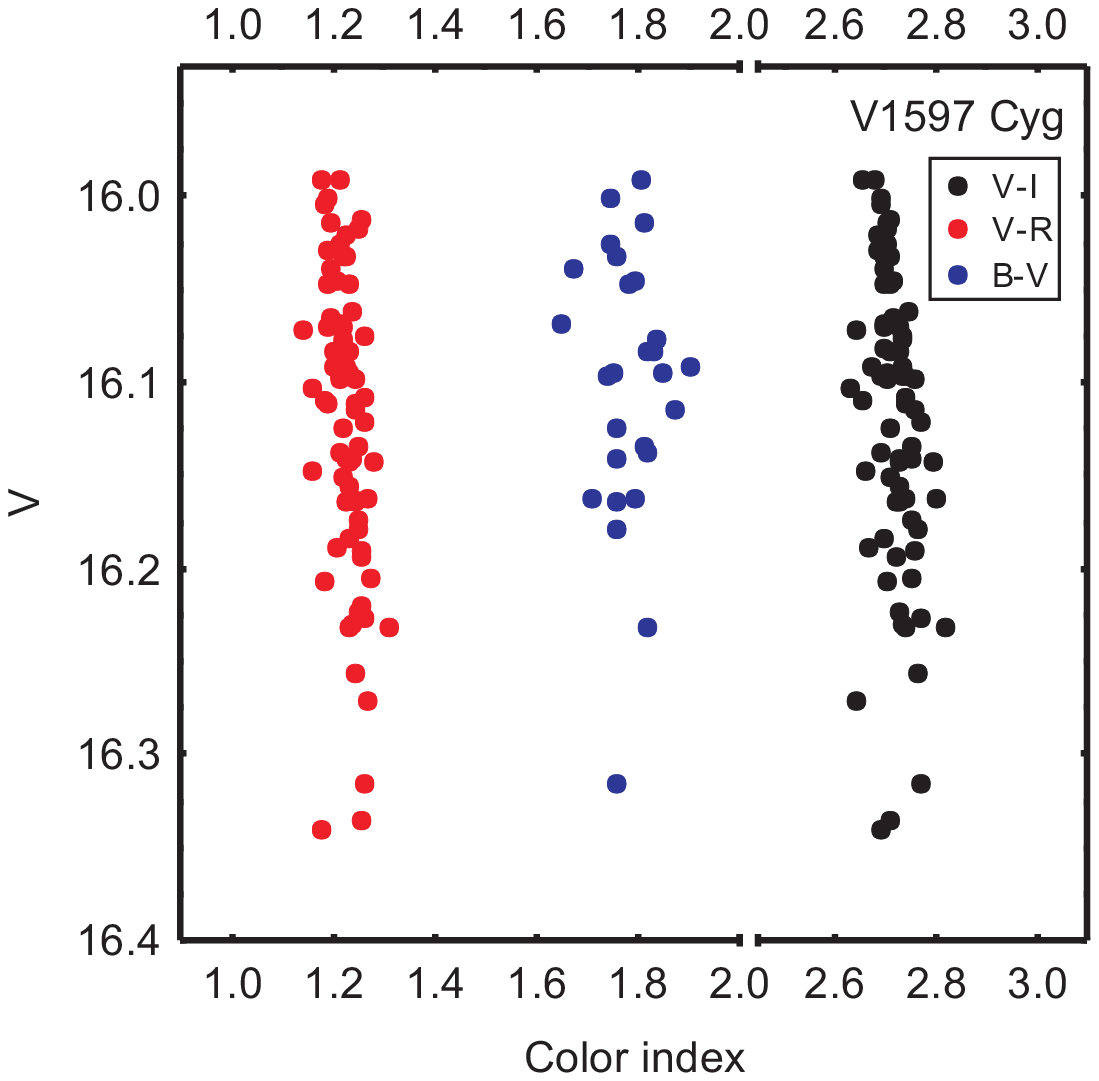}
\includegraphics[width=4.5cm]{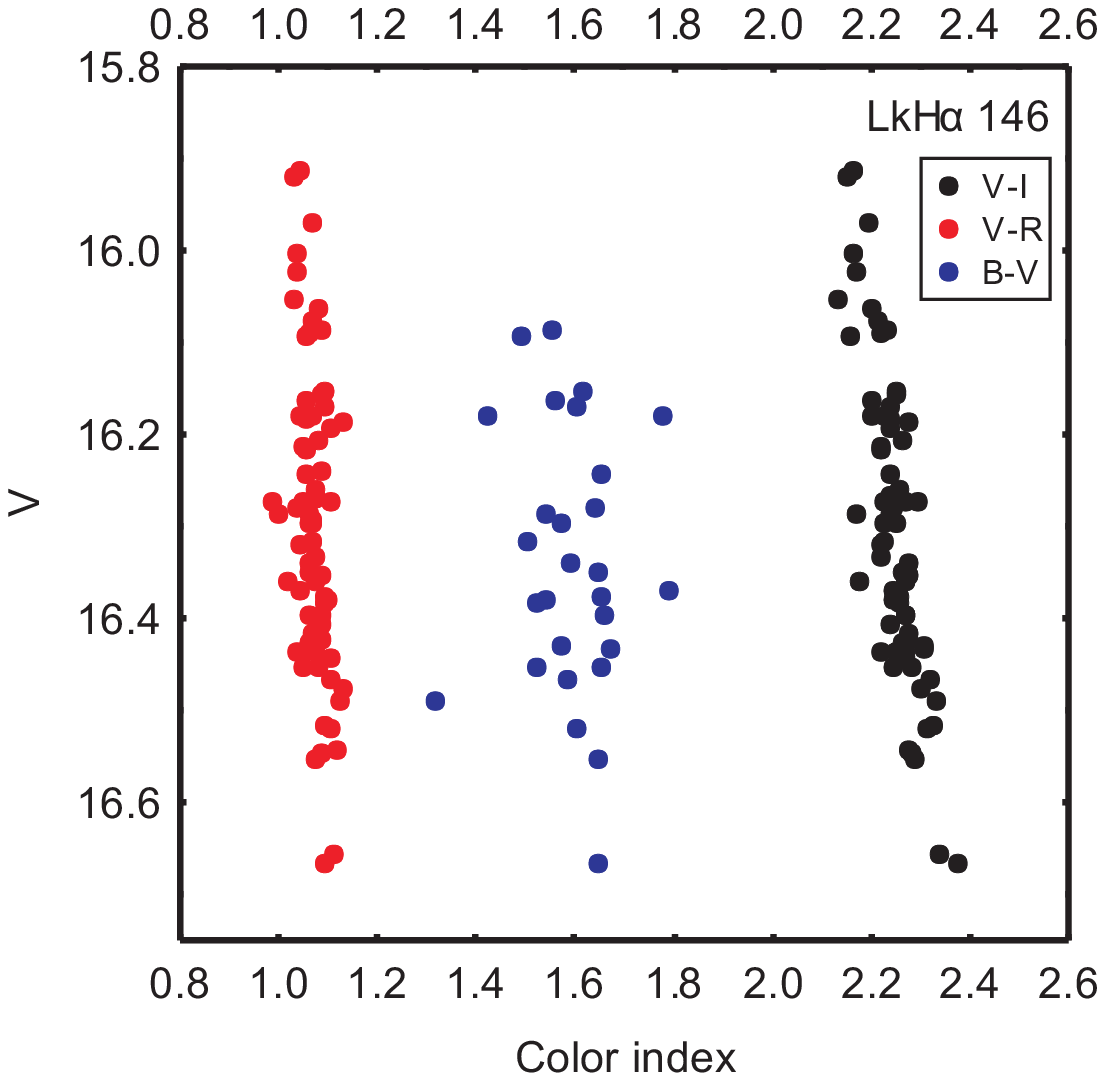}
\includegraphics[width=4.5cm]{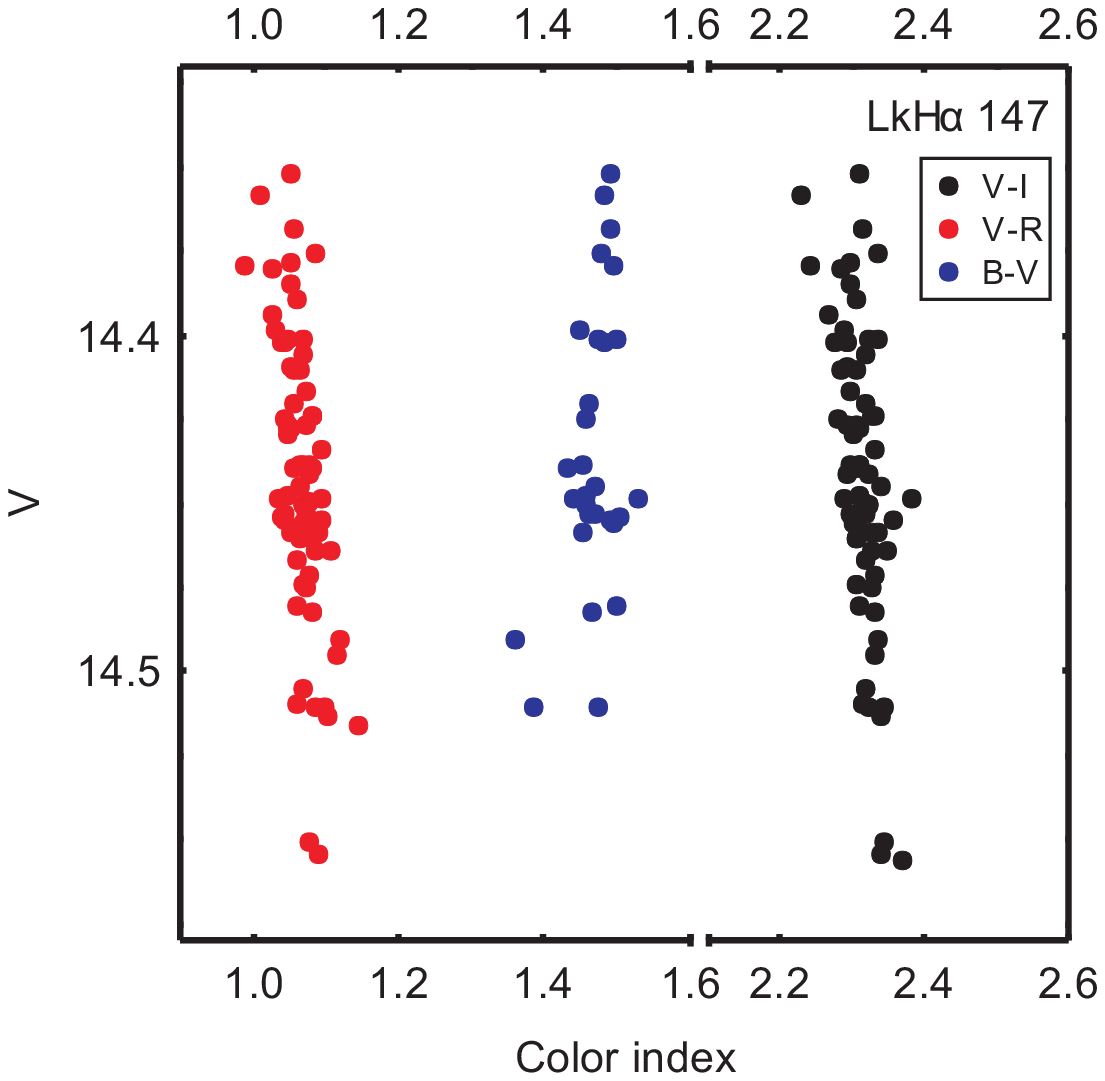}
\includegraphics[width=4.5cm]{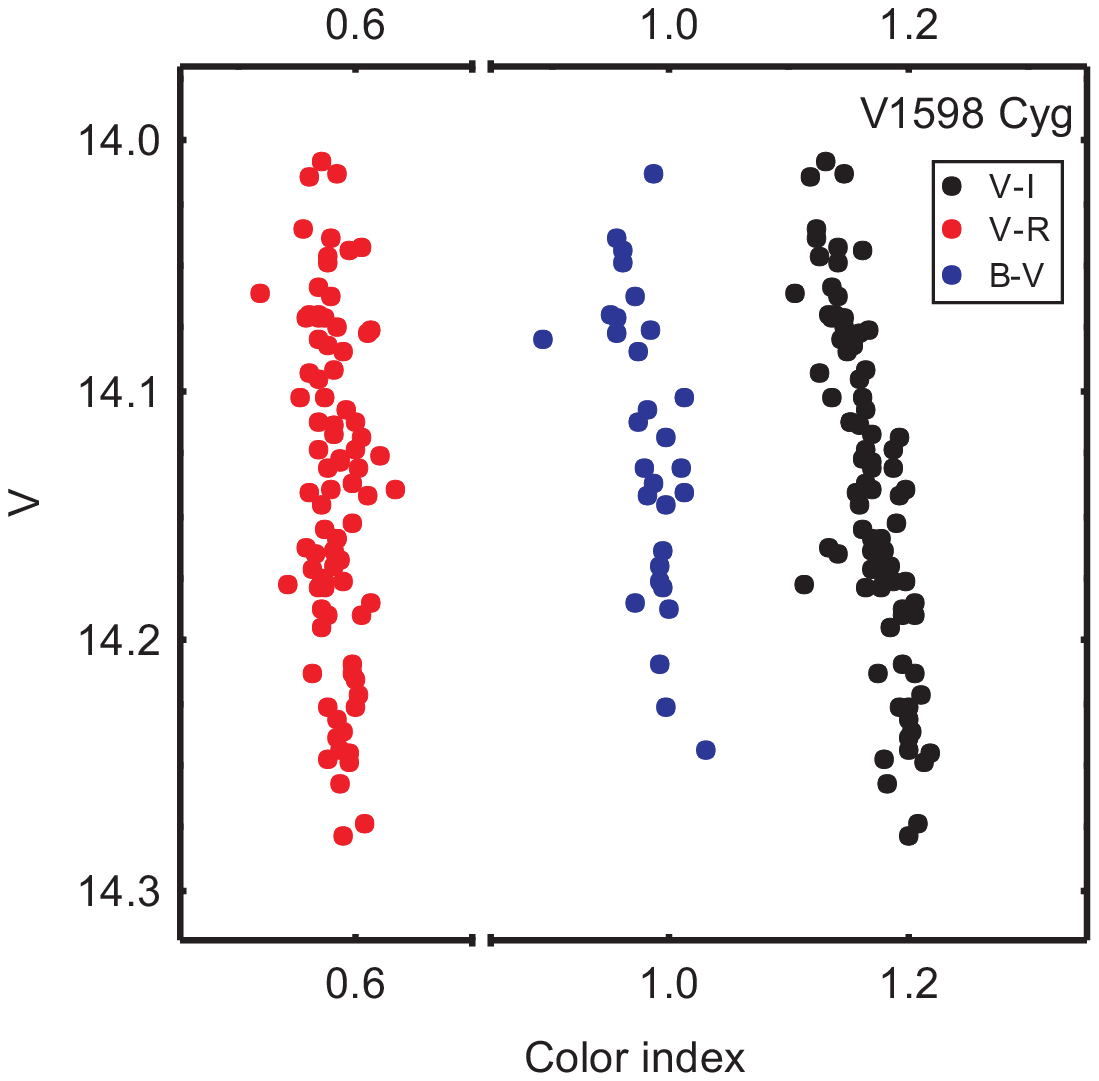}
\includegraphics[width=4.5cm]{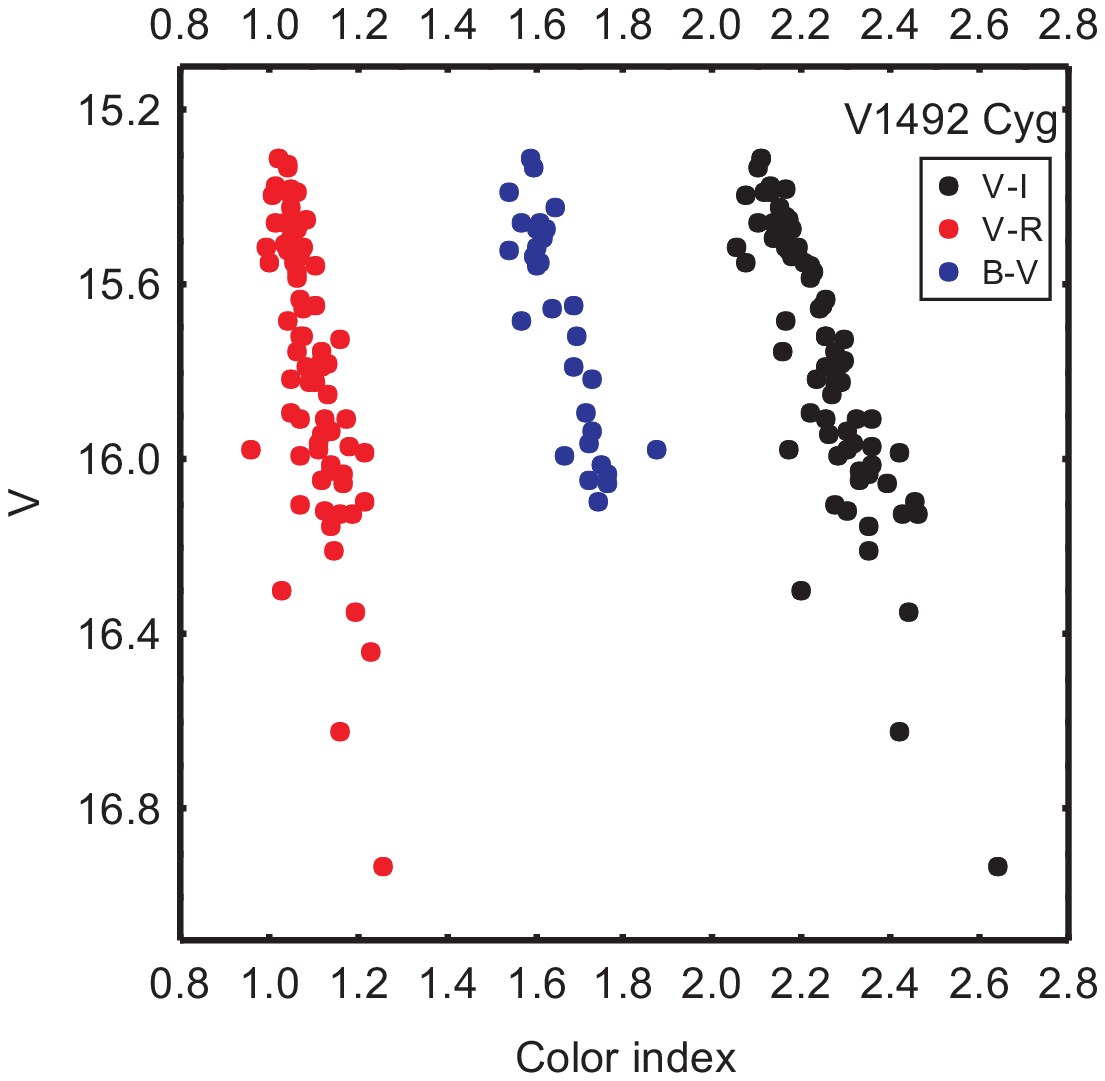}
\includegraphics[width=4.5cm]{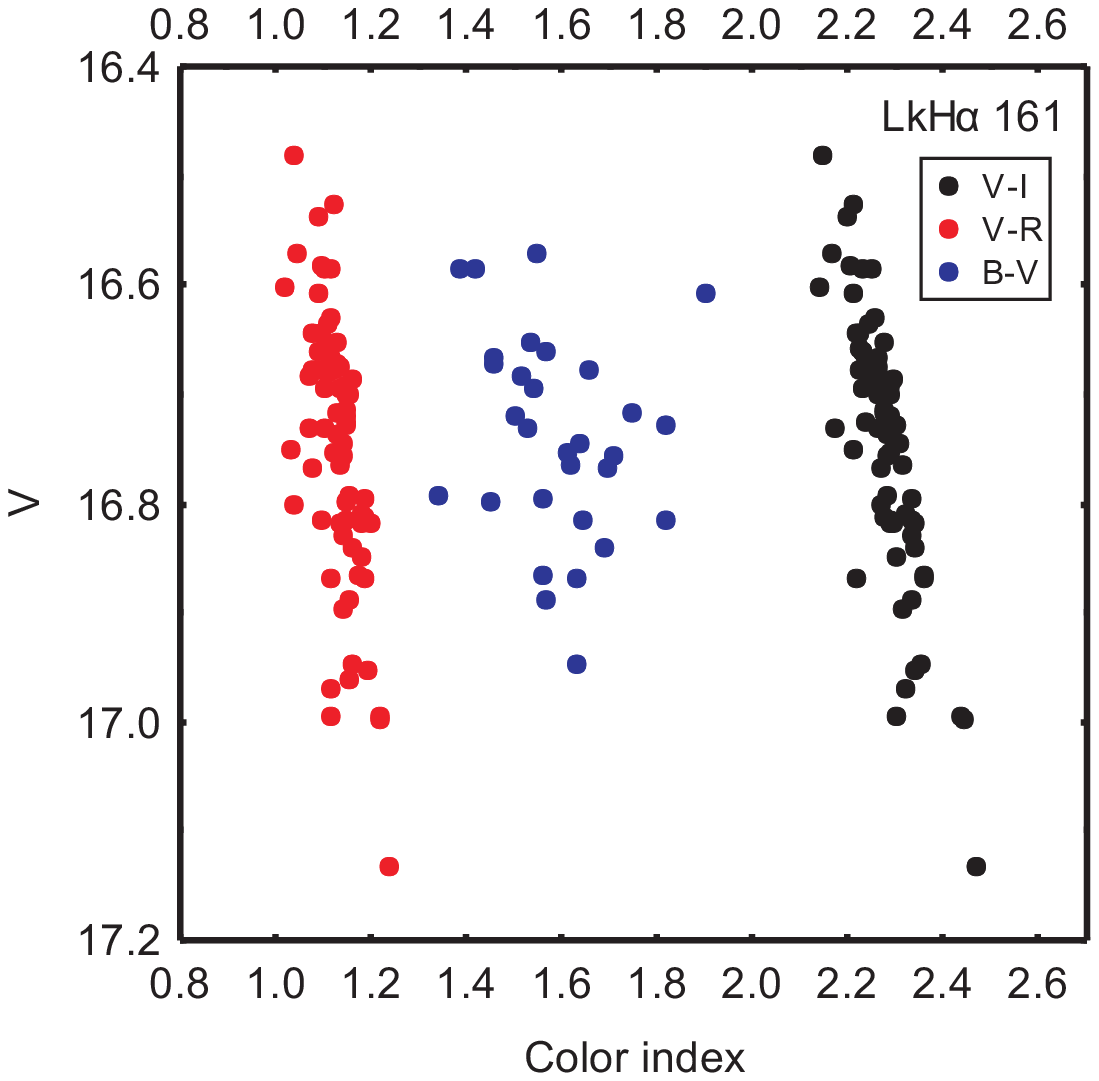}
\includegraphics[width=4.5cm]{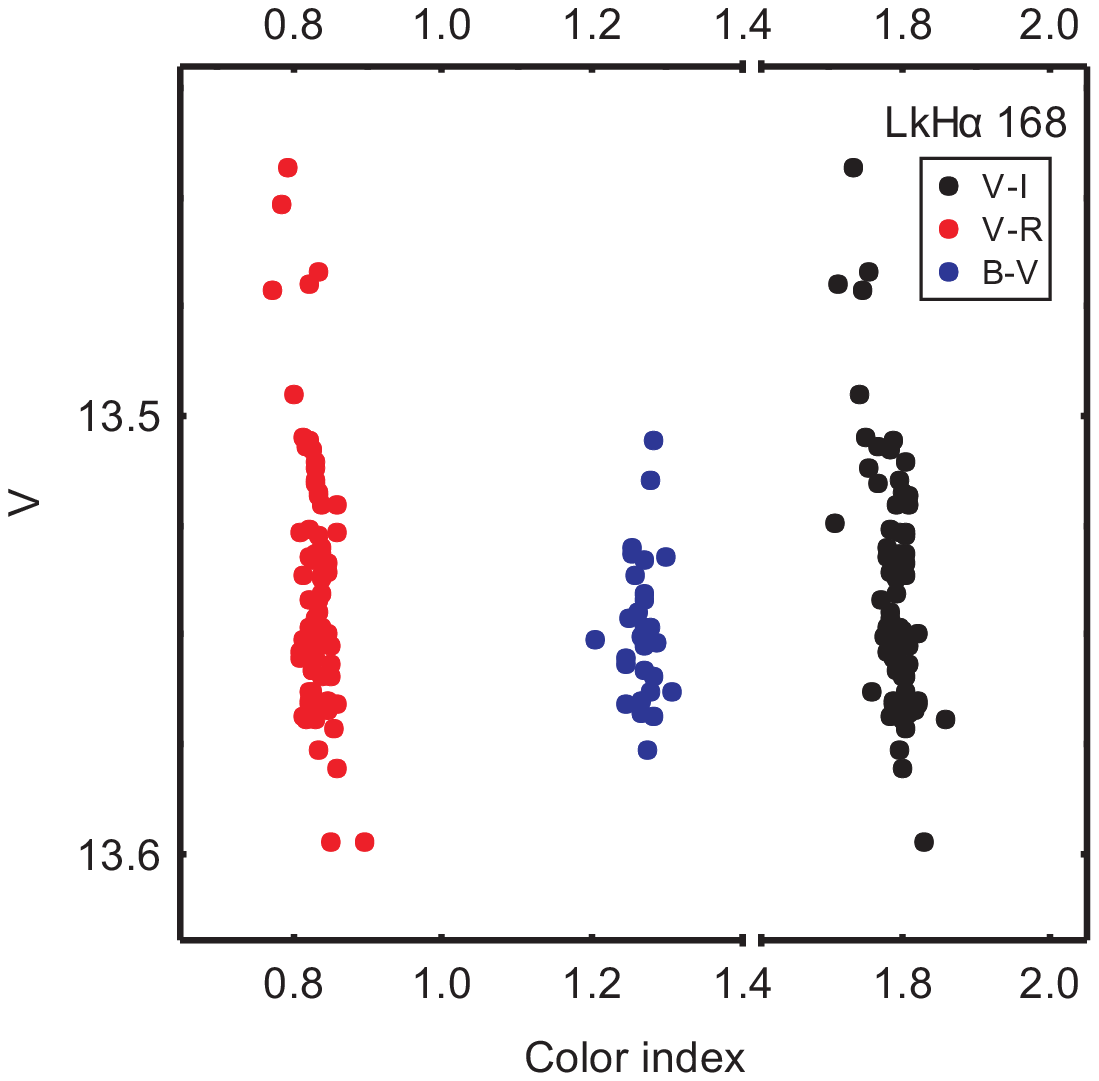}
\includegraphics[width=4.5cm]{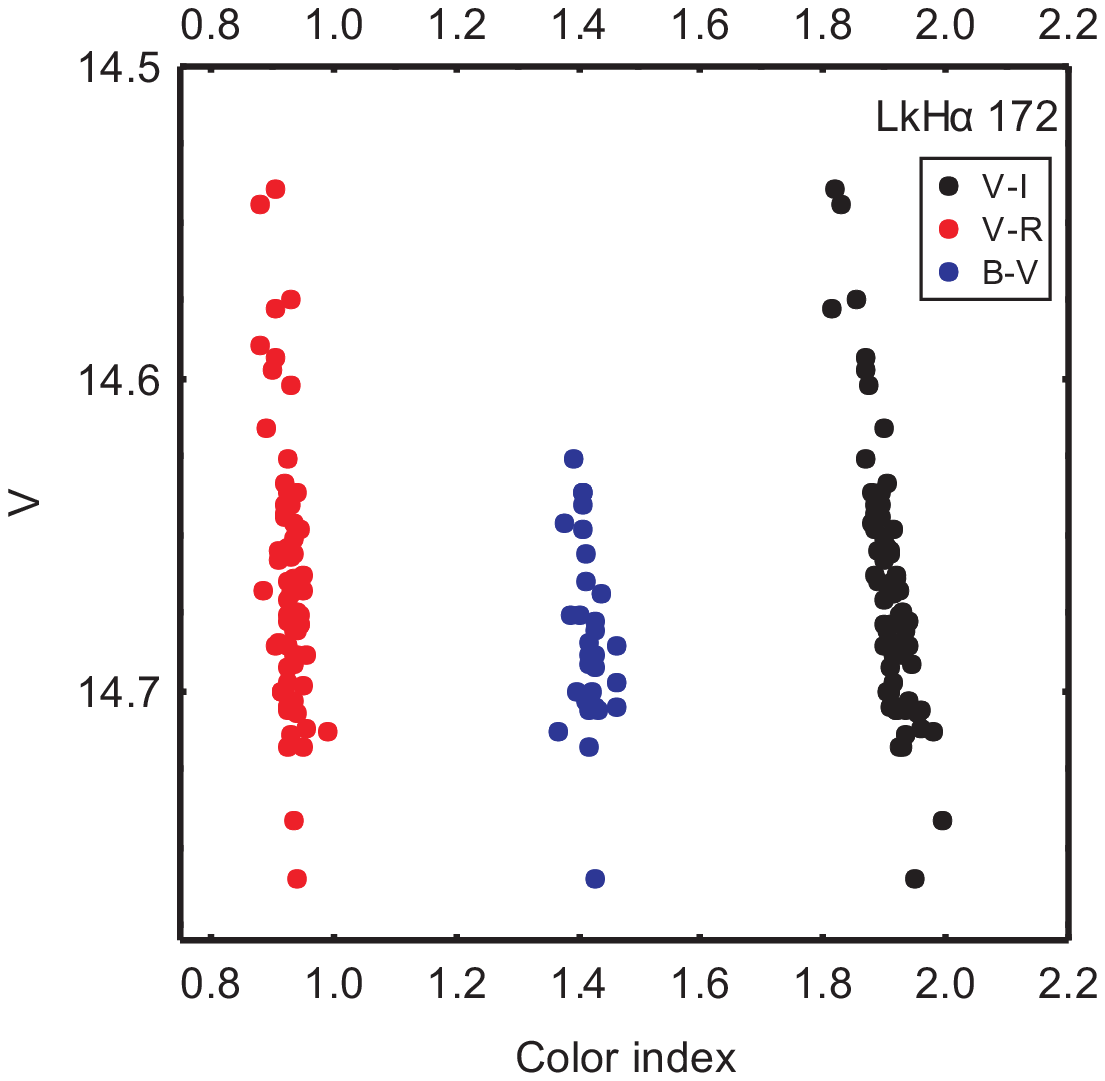}
\includegraphics[width=4.5cm]{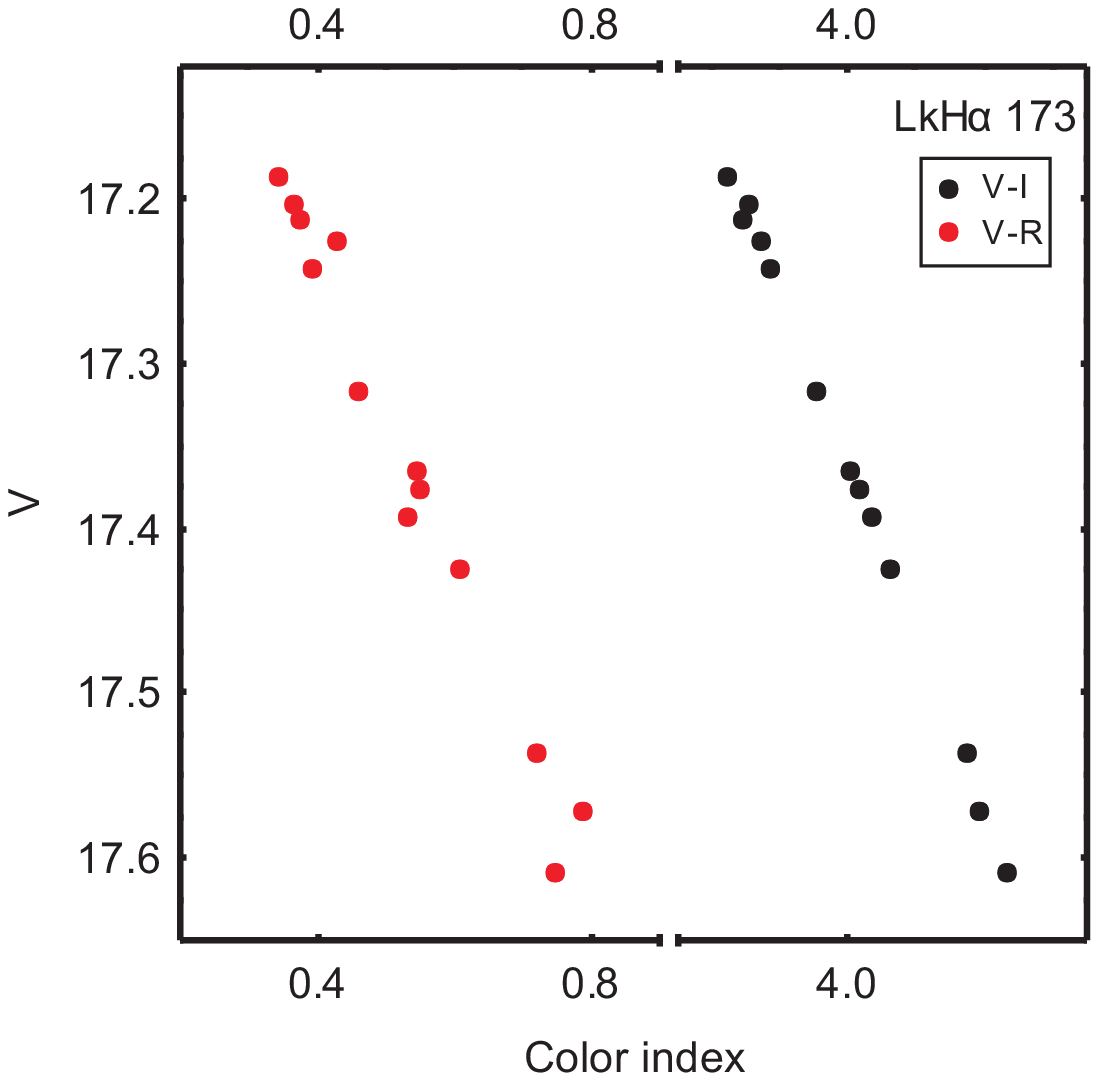}
\caption{Color indices ($V-I$, $V-R$ and $B-V$) versus the stellar $V$-magnitude of the stars from our study.}\label{Fig:colors}
\end{center}
\end{figure}

\end{document}